\documentclass[vecphys]{svmult}
\usepackage{graphicx}        
\usepackage{multicol}        
\usepackage[bottom]{footmisc}
\usepackage{amsmath,amsfonts}

\begin{document}

\DeclareGraphicsExtensions {.jpg,.pdf}

\def\msun{${\rm M_{\odot}} \;$}
\def\be{\begin{equation}}
\def\ee{\end{equation}}
\def\gcc{g cm$^{-3}$}
\def\cc{cm$^{-3}$}
\def\bi{\begin{itemize}}
\def\ei{\end{itemize}}
\newcommand{\ii}{\item}
\def\ben{\begin{enumerate}}
\def\een{\end{enumerate}}

\def\i{\item}
\def\bea{\begin{eqnarray}}
\def\eea{\end{eqnarray}}
\def\p{\partial}
\def\dt{\Delta t}

\title*{Special-relativistic Smoothed Particle Hydrodynamics: a benchmark suite}

\author{Stephan Rosswog\inst{1}}
\institute{Jacobs University Bremen, Campus Ring 1, D-28759 Bremen
\texttt{s.rosswog@jacobs-university.de}}
\maketitle

\begin{abstract}
In this paper we test a special-relativistic formulation of Smoothed Particle
Hydrodynamics (SPH) that has been derived from the Lagrangian of an ideal fluid. 
Apart from its symmetry in the particle indices, the new formulation
differs from earlier approaches in its artificial viscosity and in the use of
special-relativistic ``grad-h-terms''. In this paper we benchmark the
scheme in a number of demanding test problems. Maybe not too surprising 
for such a Lagrangian scheme, it performs close to perfectly in pure advection tests.
What is more, the method produces accurate results even in highly relativistic shock problems.
\end{abstract}


\section{Introduction}
\label{Rosswog::sec:1}
Relativity is a crucial ingredient in a variety of astrophysical phenomena. For example
the jets that are expelled from the cores of active galaxies reach velocities tantalizingly 
close to the speed of light, and motion near a black hole is heavily influenced by
space-time curvature effects.
In the recent past, substantial progress has been made in the development of numerical
tools to tackle relativistic gas dynamics problems, both on the special- and the
general-relativistic side, for reviews see \cite{Rosswog:marti03,Rosswog:font00,Rosswog:baumgarte03}. 
Most work on numerical relativistic gas dynamics has been performed in an Eulerian framework,
a couple of Lagrangian smooth particle hydrodynamics (SPH) approaches do exist though.\\
In astrophysics, the SPH method has been very successful, mainly because of its excellent conservation 
properties, its natural flexibility and robustness. Moreover, its physically intuitive 
formulation has enabled the inclusion of various physical processes beyond gas dynamics
so that many challenging multi-physics problems could be tackled. For recent reviews of 
the method we refer to the literature \cite{Rosswog:monaghan05,Rosswog:rosswog09b}. 
Relativistic versions of the SPH method were first applied to special relativity and
to gas flows evolving in a fixed background metric 
\cite{Rosswog:kheyfets90,Rosswog:mann91,Rosswog:mann93,Rosswog:laguna93a,Rosswog:chow97,Rosswog:siegler00a}. More 
recently, SPH has also been used in combination with approximative schemes to dynamically  evolve space-time 
\cite{Rosswog:ayal01,Rosswog:faber00,Rosswog:faber01,Rosswog:faber02b,Rosswog:oechslin02,Rosswog:faber04,Rosswog:faber06,Rosswog:bauswein10}.\\
In this paper we briefly summarize the main equations of a new, special-relativistic
SPH formulation that has been derived from the Lagrangian of an ideal fluid. Since 
the details of the derivation have been outlined elsewhere, we focus here on a set of numerical 
benchmark tests that complement those shown in the original paper \cite{Rosswog:rosswog09d}. 
Some of them are ``standard'' and often used to demonstrate or compare code 
performance, but most of them are more violent---and therefore more challenging---versions 
of widespread test problems.

\section{Relativistic SPH equations from a variational principle}
An elegant approach to derive relativistic SPH equations based on the discretized Lagrangian of a 
perfect fluid was suggested in \cite{Rosswog:monaghan01}. We have recently extended this approach \cite{Rosswog:rosswog09d,Rosswog:rosswog10}
by including the relativistic generalizations of what are called ``grad-h-terms'' 
in non-relativistic SPH \cite{Rosswog:springel02,Rosswog:monaghan02}. For details of the derivation we refer to the original
paper \cite{Rosswog:rosswog09d} and a recent review on the Smooth Particle Hydrodynamics method
 \cite{Rosswog:rosswog09b}.\\
In the following, we assume a flat space-time metric with signature (-,+,+,+) and use units in which the 
speed of light is equal to unity, $c=1$. We reserve Greek letters for space-time indices from 0...3
with 0 being the temporal component, while $i$ and $j$ refer to spatial components and SPH
particles are labeled by $a,b$ and $k$.\\
Using the Einstein sum convention the Lagrangian of a special-relativistic perfect fluid can 
be written as \cite{Rosswog:fock64} 
\be
L_{\rm pf,sr}= - \int T^{\mu\nu} U_\mu U_\nu \; dV\label{eq:fluid_Lag_SRT},
\ee
where
\be
T^{\mu\nu}= (n[1 + u(n,s)] + P)  U^\mu U^\nu + P \eta^{\mu\nu}
\ee
denotes the energy momentum tensor, $n$ is the baryon number density, $u$ is 
the thermal energy per baryon, $s$ the specific 
entropy, $P$ the pressure and $U^\mu=dx^\mu/d\tau$ is the four velocity with $\tau$ being 
proper time. All fluid quantities are measured in the local rest frame, 
energies are measured in units of the baryon rest mass energy\footnote{The appropriate mass $m_0$ obviously
  depends on the ratio of neutrons to protons, i.e. on the nuclear composition
of the considered fluid.}, $m_0 c^2$.
For practical simulations we give up general covariance and perform the calculations
in a chosen ``computing frame'' (CF). In the general case, a fluid element moves with 
respect to this frame, therefore,  the baryon number density in the CF, $N$, is related 
to the local fluid rest frame via a Lorentz contraction
\be
N= \gamma n, \label{Rosswog::eq:N_vs_n}
\ee
where $\gamma$ is  the Lorentz factor of the fluid element as measured in the CF.
The simulation volume in the CF can be subdivided into volume elements
such that each element $b$ contains $\nu_b$ baryons and these volume elements, 
$\Delta V_b= \nu_b/N_b$, can be used in the SPH discretization process of a 
quantity $f$:
\be
f(\vec{r})= \sum_b f_b \frac{\nu_b}{N_b} W(|\vec{r}-\vec{r}_b|,h),\label{eq:SPH_discret}
\ee
where the index labels quantities at the position of particle $b$, $\vec{r}_b$.
Our notation does not distinguish between the approximated values (the
$f$ on the LHS) and the values at the particle positions ($f_b$ on the
RHS). The quantity $h$ is the smoothing length that characterizes the width
of the smoothing kernel $W$, for which we apply the cubic spline kernel that is
commonly used in SPH \cite{Rosswog:monaghan92,Rosswog:monaghan05}. Applied to the baryon number density 
in the CF at the position of particle $a$, Eq.~(\ref{eq:SPH_discret}) yields:
\be
N_a= N(\vec{r}_a)= \sum_b \nu_b W(|\vec{r}_a-\vec{r}_b|,h_a).\label{eq:dens_summ_SR}
\ee
This equation takes over the role of the usual density summation of
non-relativistic SPH, $\rho (\vec{r}_a)= \sum_b m_b
W(|\vec{r}_a-\vec{r}_b|,h)$. Since we keep the baryon numbers associated with each
SPH particle, $\nu_b$, fixed, there is no need to evolve a continuity equation
and baryon number is conserved by construction. If desired, the continuity equation
can be solved though, see e.g. \cite{Rosswog:chow97}. Note that we have used $a$'s own smoothing 
length in evaluating the kernel in Eq.~(\ref{eq:dens_summ_SR}). To fully exploit the natural 
adaptivity of a particle method, we adapt the smoothing length according to 
\be
h_a= \eta \left(\frac{\nu_a}{N_a}\right)^{-1/D}\label{eq:dens_summ_SR_N_b},
\ee
where $\eta$ is a suitably chosen numerical constant, usually in the range between 1.3 and 1.5,
and $D$ is the number of spatial dimensions. 
Hence, similar to the non-relativistic case \cite{Rosswog:springel02,Rosswog:monaghan02}, the density and the
smoothing length mutually depend on each other and a self-consistent solution for both can be obtained
by performing an iteration until convergence is reached.\\
With these prerequisites at hand, the fluid Lagrangian can be discretized \cite{Rosswog:monaghan01,Rosswog:rosswog09b}
\be
L_{\rm SPH,sr}= - \sum_b \frac{\nu_b}{\gamma_b} [1+ u(n_b,s_b)].\label{eq:SR:L_SPH}
\ee
Using the first law of thermodynamics one finds (for a detailed derivation see Sec. 4 in \cite{Rosswog:rosswog09b})
for the canonical momentum per baryon
\bea
\vec{S}_a \equiv \frac{1}{\nu_a} \frac{\partial L_{\rm SPH,sr}}{\partial\vec{v}_a}
= \gamma_a \vec{v}_a \left(1+u_a+\frac{P_a}{n_a}\right)
\label{eq:can_mom},
\eea
which is the quantity that we evolve numerically.
Its evolution equation follows from the Euler-Lagrange equations,
\be
\frac{d}{dt} \frac{\p L}{\p \vec{v}_a} - \frac{\p L}{\p \vec{r}_a}= 0,
\ee
 as \cite{Rosswog:rosswog09b}
\be
\frac{d\vec{S}_a}{dt}= - \sum_b \nu_b \left(
\frac{P_a}{N_a^2 \Omega_a} \nabla_a W_{ab}(h_a) + 
\frac{P_b}{N_b^2 \Omega_b} \nabla_a W_{ab}(h_b)
\right),\label{eq:momentum_eq_no_diss}
\ee
where the ``grad-h'' correction factor
\be
\Omega_b\equiv 1-\frac{\p h_b}{\p N_b} \sum_k \frac{\p W_{bk}(h_b)}{\p h_b}
\ee
was introduced. 
As numerical energy variable we use the canonical energy per baryon,
\be
\epsilon_a \equiv \gamma_a \left(1+u_a+\frac{P_a}{n_a}\right) -
\frac{P_a}{N_a}= \vec{v}_a \cdot \vec{S}_a + \frac{1+u_a}{\gamma_a}\label{eq:SR:epsilon_a}
\ee
which evolves according to \cite{Rosswog:rosswog09b}
\be
\frac{d \epsilon_a}{dt} = - \sum_b \nu_b 
\left(
\frac{P_a \vec{v}_b}{N_a^2 \Omega_a} \cdot \nabla_a W_{ab}(h_a) + 
\frac{P_b \vec{v}_a}{N_b^2 \Omega_b} \cdot \nabla_a W_{ab}(h_b)
\right).
\label{eq:ener_eq_no_diss}
\ee 
As in grid-based approaches, at each time step a conversion between the numerical and the physical
variables is required \cite{Rosswog:chow97,Rosswog:rosswog09d}.\\
The set of equations needs to be closed by an equation of state. In all of the following tests, we use a
polytropic equation of state, $P= (\Gamma-1)n u$, where $\Gamma$ is the polytropic exponent (keep in mind
our convention of measuring energies in units of $m_0 c^2$).

\section{Artificial dissipation}

To handle shocks, additional artificial dissipation terms need to be included.
We use terms similar to \cite{Rosswog:chow97}
\be
\left(\frac{d\vec{S}_a}{dt}\right)_{\rm diss}= - \sum_b \nu_b \Pi_{ab} 
\overline{\nabla_a W_{ab}} \quad {\rm with} \quad \Pi_{ab}= - 
\frac{K v_{\rm sig}}{\bar{N}_{ab}} (\vec{S}_a^\ast-\vec{S}_b^\ast) \cdot\hat{e}_{ab}
\label{Rosswog::eq:diss_mom}
\ee
and 
\be
\left(\frac{d\epsilon_a}{dt}\right)_{\rm diss}=  - \sum_b \nu_b \vec{\Psi}_{ab} \cdot
\overline{\nabla_a W_{ab}} \quad {\rm with} \quad \vec{\Psi}_{ab} = -
\frac{K v_{\rm sig}}{\bar{N}_{ab}} (\epsilon_a^\ast-\epsilon_b^\ast)\hat{e}_{ab}. 
\label{Rosswog::eq:diss_en}
\ee
Here $K$ is a numerical constant of order unity, $v_{\rm sig}$ an
appropriately chosen signal velocity, see below, $\bar{N}_{ab}= (N_a+N_b)/2$, and
$\hat{e}_{ab}= (\vec{r}_a-\vec{r}_b) / |\vec{r}_a-\vec{r}_b|$
is the unit vector pointing from particle $b$ to particle $a$. For the symmetrized kernel gradient
we use  
\be
\overline{\nabla_a W_{ab}} = \frac{1}{2}\left[\nabla_a W_{ab}(h_a) +  \nabla_a W_{ab}(h_b) \right].
\ee
Note that in \cite{Rosswog:chow97} $\nabla_a W_{ab}(h_{ab})$ was used instead of our $\overline{\nabla_a W_{ab}}$,
in practice we find the differences between the two symmetrizations negligible.
The stars at the variables in Eqs.~(\ref{Rosswog::eq:diss_mom}) and (\ref{Rosswog::eq:diss_en}) indicate that the projected 
Lorentz factors
\be
\gamma_k^\ast= \frac{1}{\sqrt{1-(\vec{v}_k\cdot \hat{e}_{ab})^2}}
\ee
are used instead of the normal Lorentz factor. 
This projection onto the line connecting particle $a$ and $b$ has been chosen to guarantee 
that the viscous dissipation is positive definite \cite{Rosswog:chow97}.\\
The signal velocity, $v_{\rm sig}$, is an
estimate for the speed of approach of a signal sent from particle $a$ 
to particle $b$. The idea is to have a robust estimate that does 
not require much computational effort.  We use \cite{Rosswog:rosswog09d} 
\be
v_{\rm sig,ab}= {\rm max}(\alpha_a,\alpha_b),\label{eq:vsig}
\ee
where
\be
\alpha_k^{\pm}= {\rm max}(0,\pm \lambda^\pm_k)
\ee
with $\lambda^\pm_k$ being the extreme local eigenvalues of the Euler equations
\be
\lambda^\pm_k= \frac{v_k\pm c_{{\rm s},k}}{1\pm v_k c_{{\rm s},k}}
\ee
and $c_{{\rm s},k}$ being the relativistic sound velocity of particle $k$.
These 1D estimates can be generalized to higher spatial dimensions, see e.g. \cite{Rosswog:marti03}.
The results are not particularly sensitive to the exact form of the signal velocity, but
in experiments we find that
Eq.~(\ref{eq:vsig}) yields somewhat crisper shock fronts and less smeared contact 
discontinuities (for the same value of $K$) than earlier suggestions \cite{Rosswog:chow97}.\\
Since we are aiming at solving the relativistic evolution equations of an {\em ideal} fluid, we
want dissipation only where it is really needed, i.e. near shocks where entropy needs to be 
produced\footnote{A description of the general reasoning behind artificial viscosity can be found, for example, 
in Sec. 2.7 of \cite{Rosswog:rosswog09b}}. To this end, we assign an individual value of the 
parameter $K$ to each SPH particle and integrate an additional differential equation 
to determine its value. For the details of the time-dependent viscosity parameter treatment
we refer to \cite{Rosswog:rosswog09d}.

\section{Test bench}
In the following we demonstrate the performance of the above described scheme
at a slew of benchmark tests.
The exact solutions of the Riemann problems have been obtained by help of
the RIEMANN\_VT.f code provided by Marti and M\"uller \cite{Rosswog:marti03}.
Unless mentioned otherwise, approximately 3000 particles are shown.

\subsection{Test 1: Riemann problem 1}
\begin{figure}[htbp] 
 \centerline{\includegraphics[height=0.3\textheight]{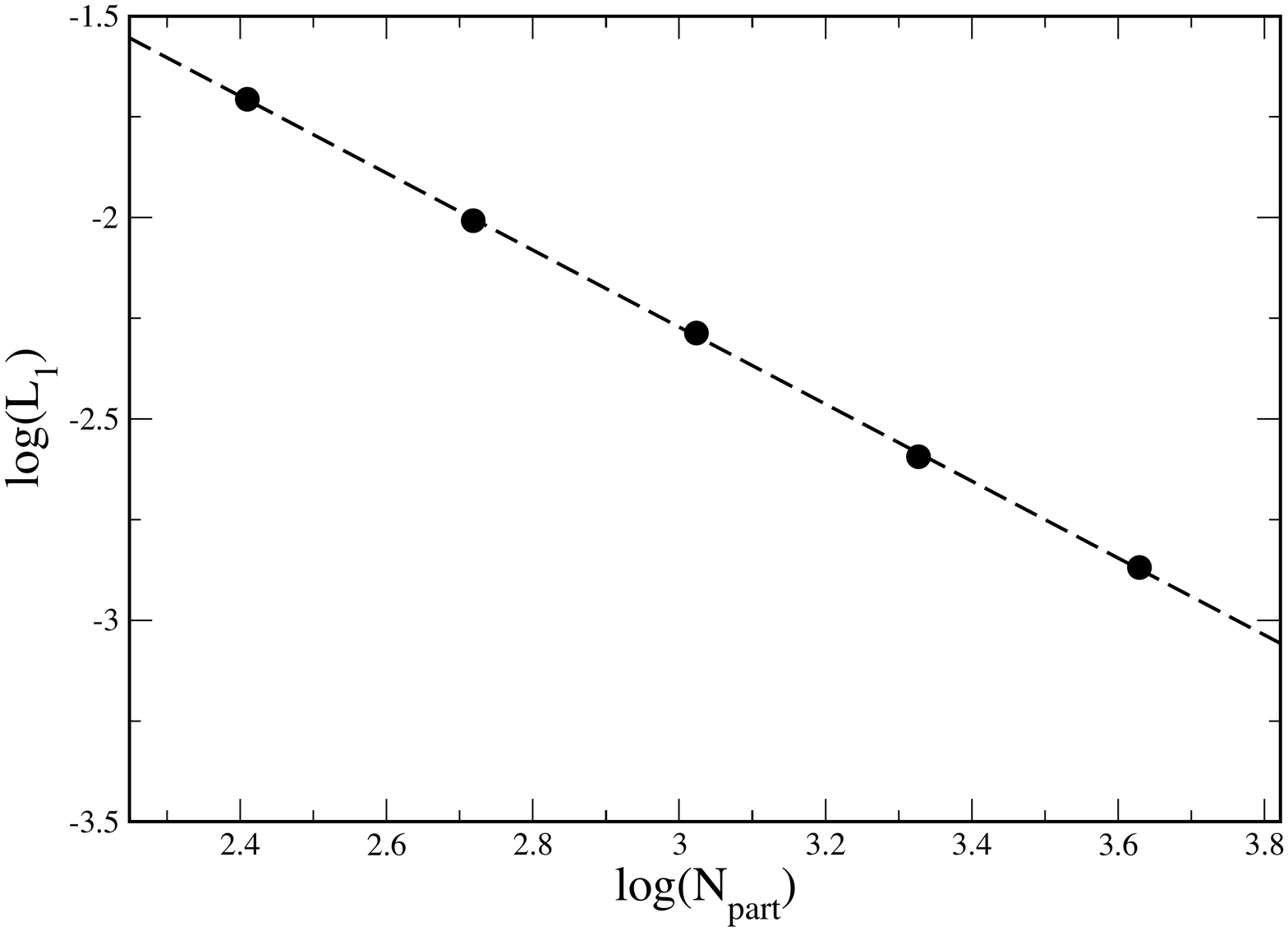}}
   \caption{Decrease of the error as defined in Eq.~(\ref{eq:L1}) as a function of particle number for the relativistic
            shock tested in Riemann problem 1. The error decreases close to $L_1 \propto N_{\rm part}^{-1}$.}
   \label{Rosswog::fig:2}
\end{figure}

This moderately relativistic (maximum Lorentz factor $\gamma_{\max}\approx 1.4$)
shock tube has become a standard touch-stone for relativistic hydrodynamics codes 
\cite{Rosswog:hawley84b,Rosswog:marti96,Rosswog:chow97,Rosswog:siegler00b,Rosswog:delZanna02,Rosswog:marti03}.
It uses a polytropic equation of state (EOS) with an exponent of $\Gamma=5/3$
and $[P, N, v]_{\rm L}= [40/3, 10, 0]$ for the left-hand state and 
$[P, N, v]_{\rm R}= [10^{-6}, 1, 0]$ for the right-hand state.
\begin{figure}[htbp] 
   \centering
   \centerline{
   \includegraphics[height=0.3\textheight]{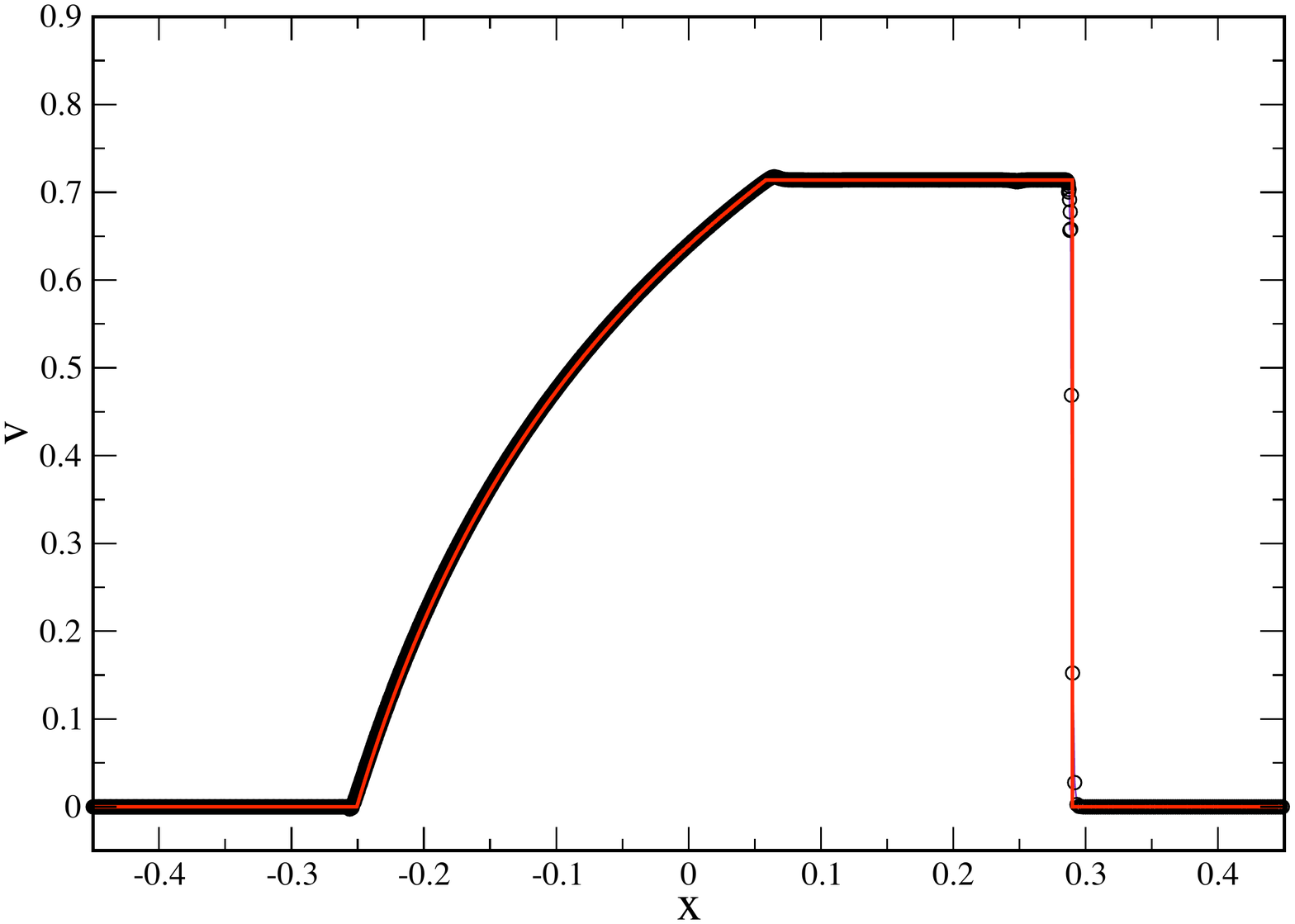} \hspace*{-0.8cm}
   \includegraphics[height=0.3\textheight]{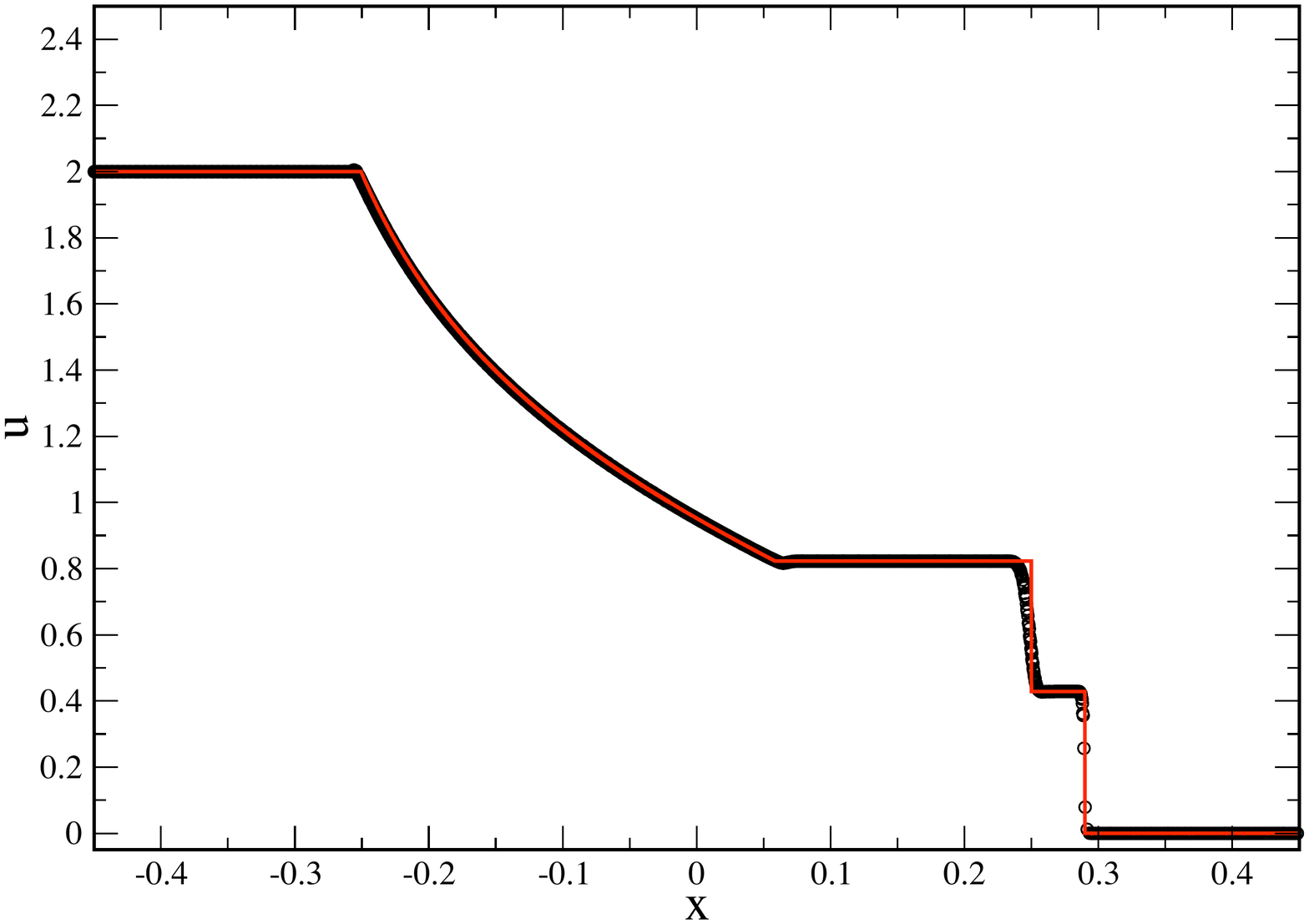}}
 \vspace*{-0.6cm}
   \centerline{
   \includegraphics[height=0.3\textheight]{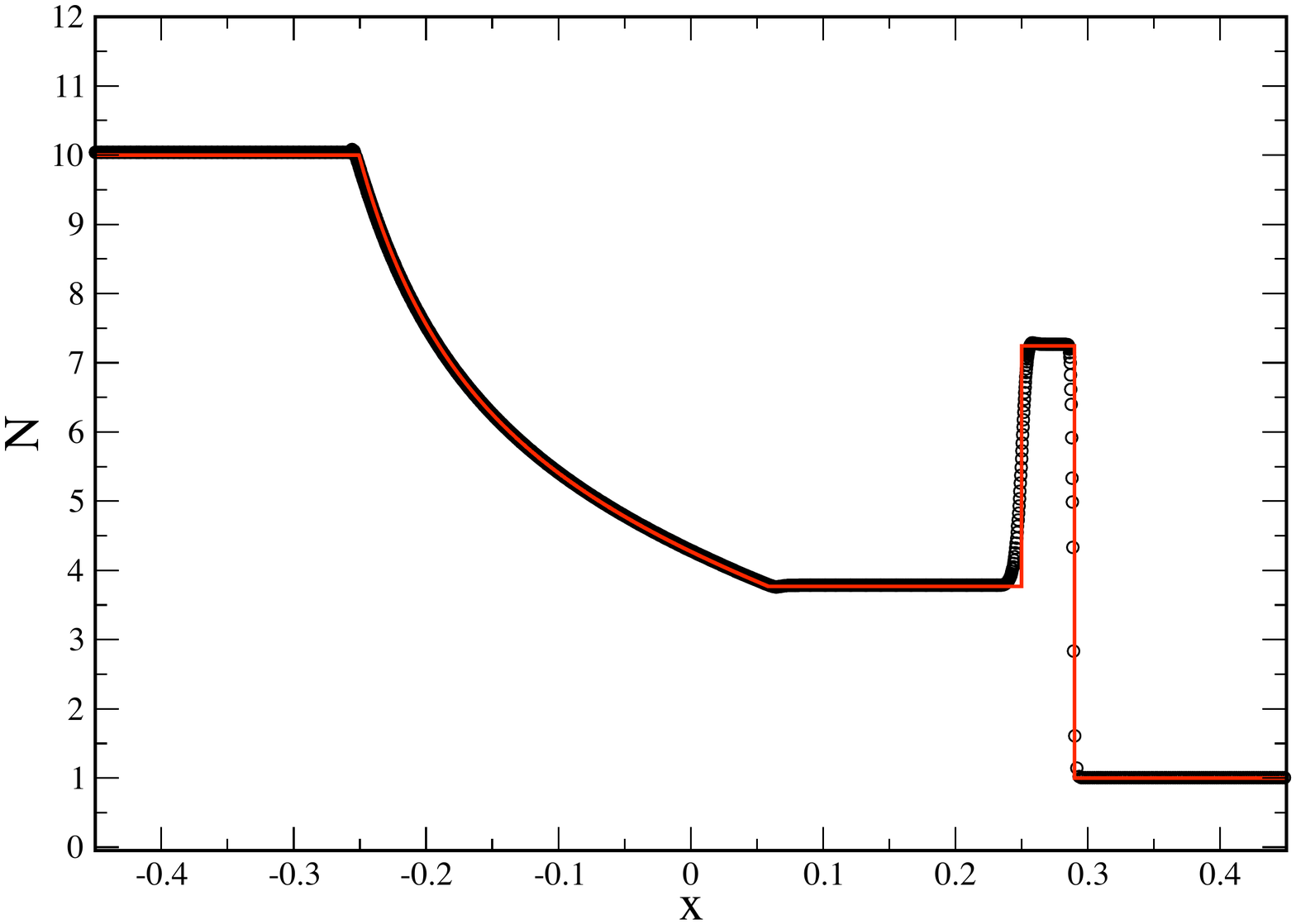} \hspace*{-0.8cm}
   \includegraphics[height=0.3\textheight]{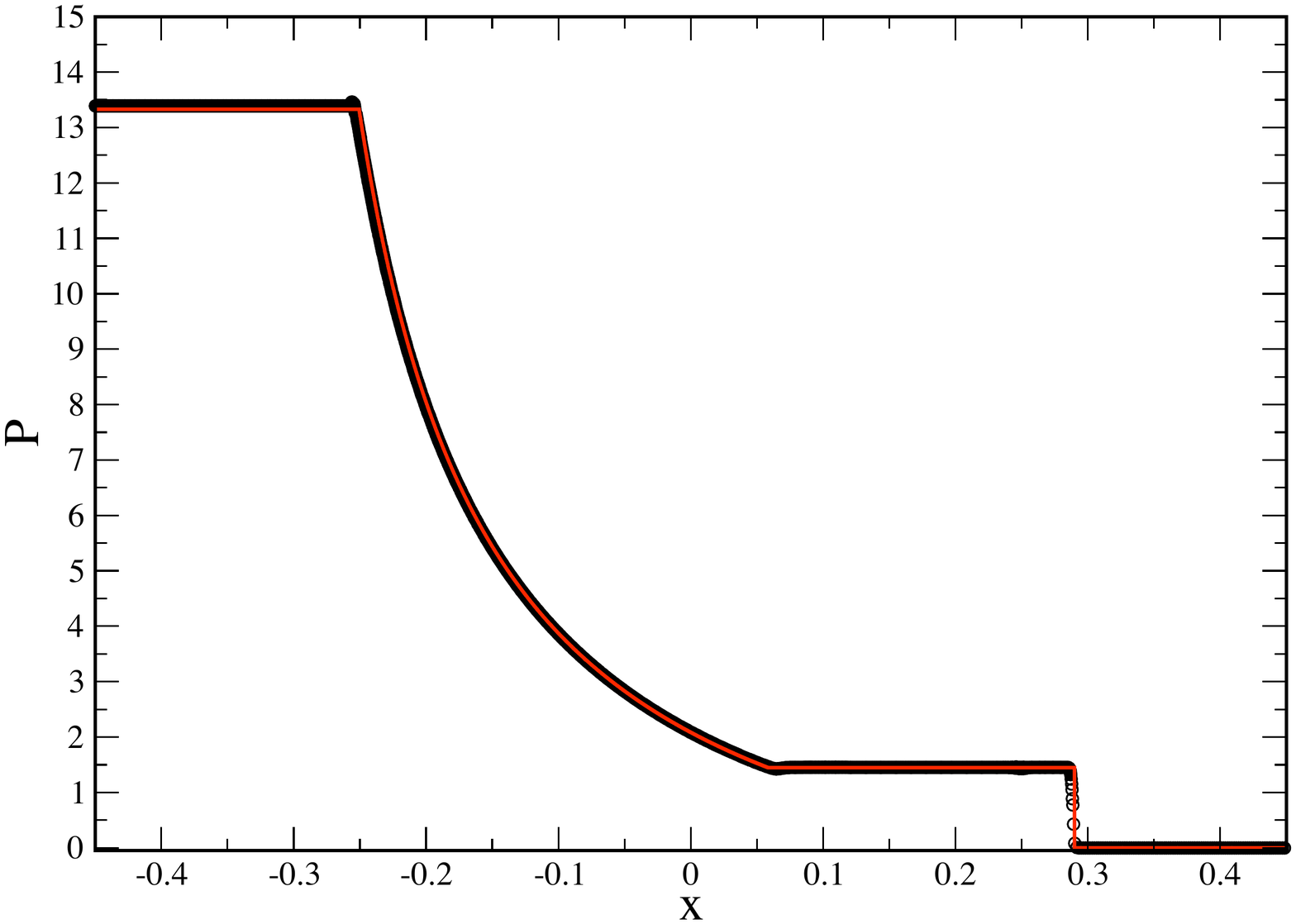}}
   \caption{Results of the relativistic shock tube of test 1 at $t=0.35$: SPH results (circles) vs. exact solution
            (red line). From left to right, top to bottom: velocity (in units of $c$), specific energy,
            computing frame baryon number density and pressure.}
   \label{Rosswog::fig:1}
\end{figure}
As shown in Fig.~\ref{Rosswog::fig:1}, the numerical solution at $t=0.35$  (circles) agrees nearly perfectly 
with the exact one. Note in particular the absence of any spikes in $u$ and $P$ at the contact discontinuity
(near $x\approx 0.25$), such spikes had plagued many earlier relativistic SPH formulations \cite{Rosswog:laguna93a,Rosswog:siegler00a}.
The only places where we see possibly room for improvement is the contact 
discontinuity  which is slightly smeared out and the slight over-/undershoots at the edges
of the rarefaction fan.\\
In order to monitor how the error in the numerical solution decreases as a function of increased resolution,
we calculate
\be
L_1\equiv \frac{1}{N_{\rm part}} \sum_b^{N_{\rm part}} |v_b - v_{\rm ex}(r_b)|,\label{eq:L1}
\ee
where $N_{\rm part}$ is the number of SPH-particles, $v_b$ the (1D) velocity of SPH-particle $b$ and 
$v_{\rm ex}(r_b)$ the exact solution for the velocity at position $r_b$. The results for $L_1$ are displayed in 
Fig.~\ref{Rosswog::fig:2}. The error  $L_1$ decreases close to $\propto N_{\rm part}^{-1}$ (actually, the best fit 
is $L_1\propto N_{\rm part}^{-0.96}$), which is what is also found for Eulerian methods in tests that involve shocks.
Therefore, for problems that involve shocks we consider the method first-order accurate.
The order of the method for smooth flows will be determined in the context of test 6. 

\subsection{Test 2: Riemann problem 2}
\begin{figure}[htbp] 
   \centering
 \centerline{
   \includegraphics[height=0.3\textheight]{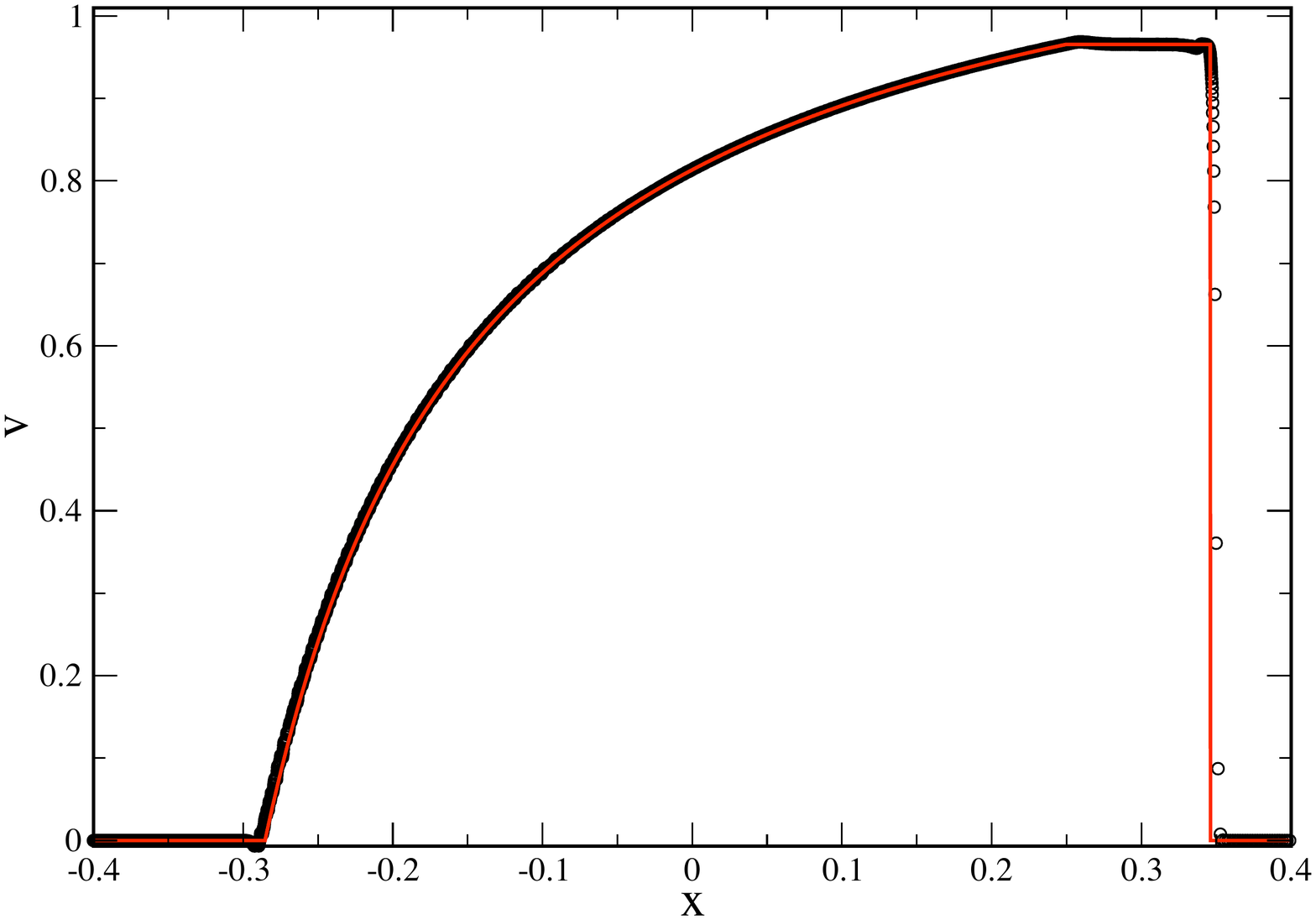} \hspace*{-0.8cm}
   \includegraphics[height=0.3\textheight]{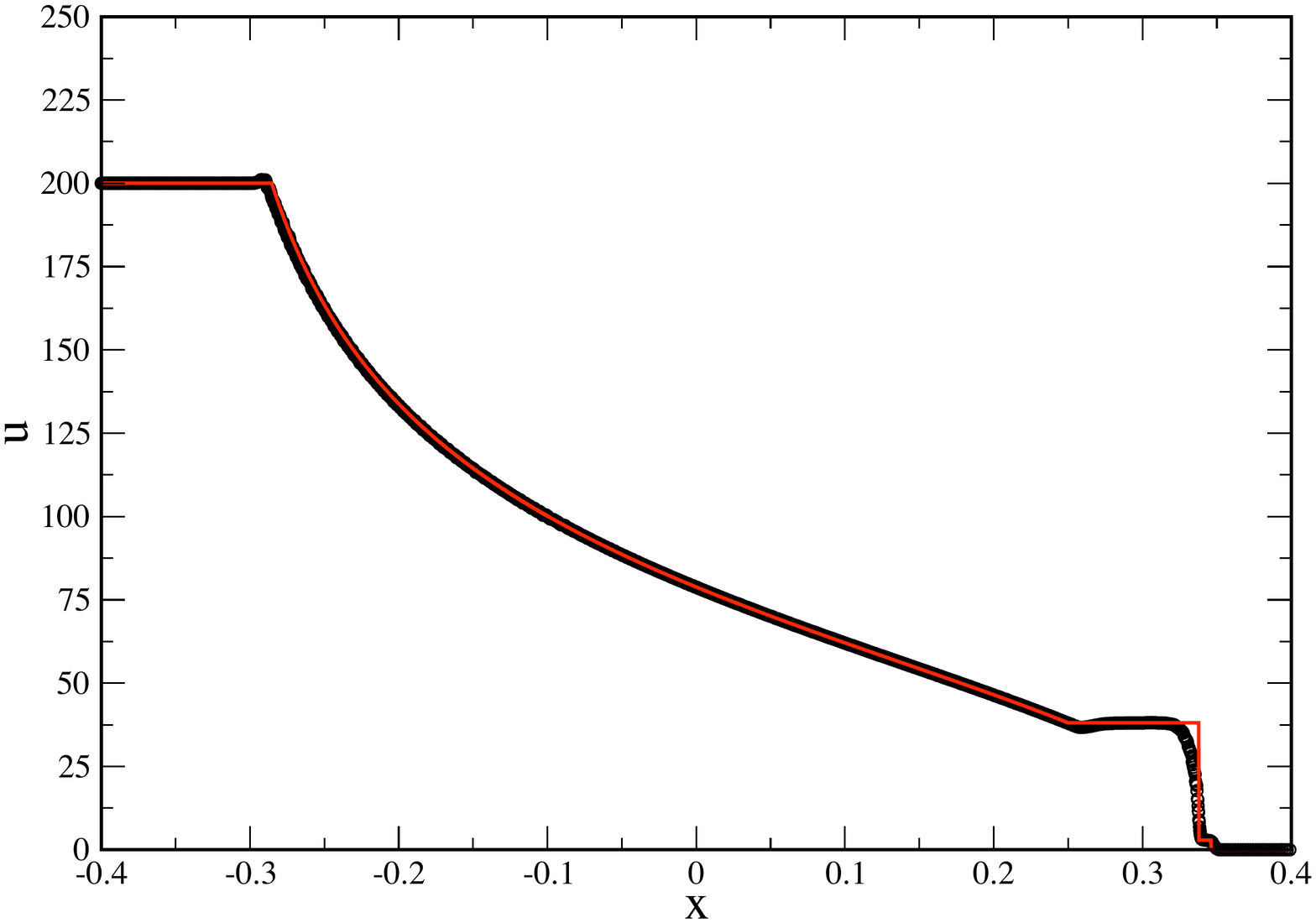}}
 \vspace*{-0.6cm}
   \centerline{
   \includegraphics[height=0.3\textheight]{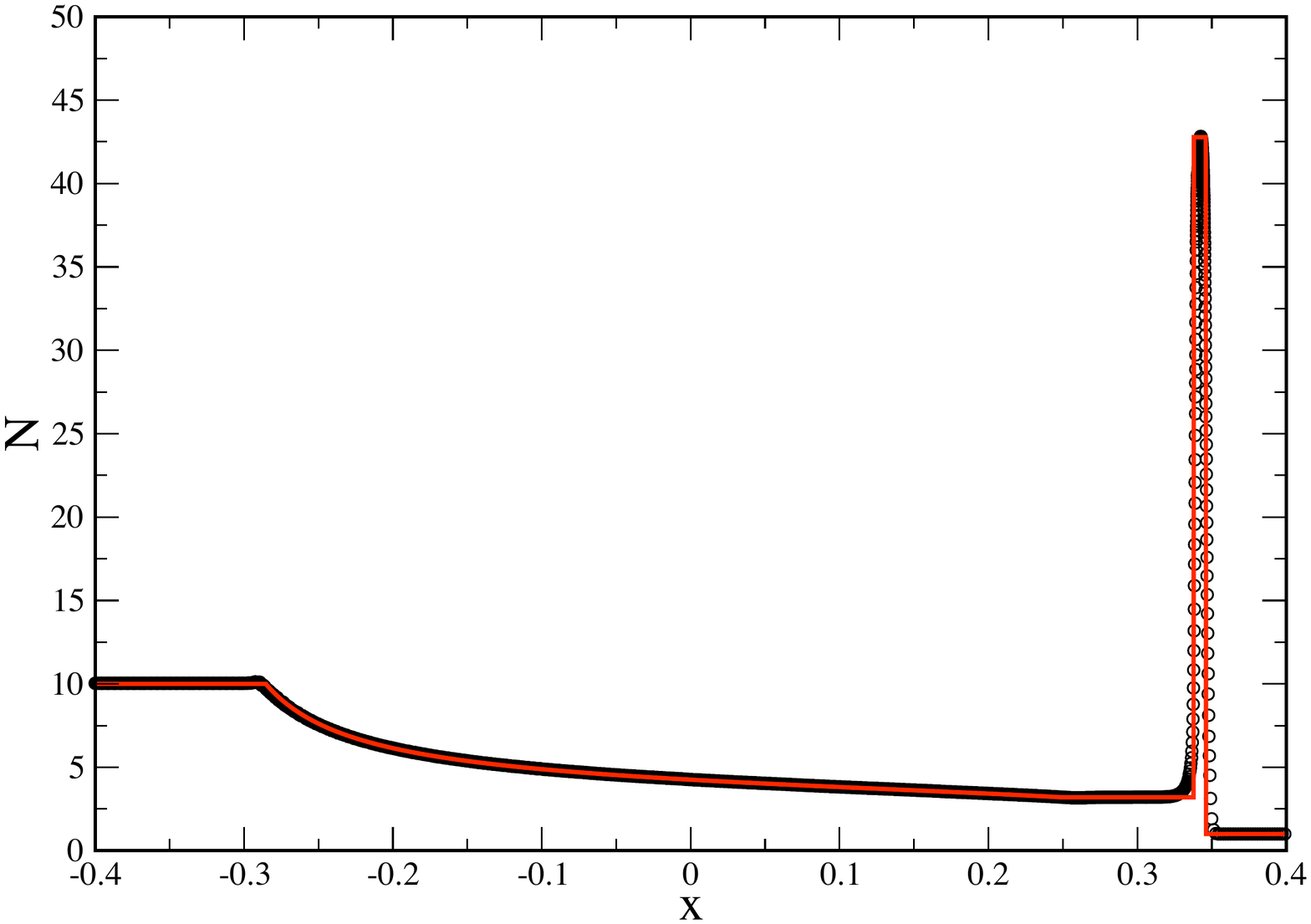} \hspace*{-0.8cm}
   \includegraphics[height=0.3\textheight]{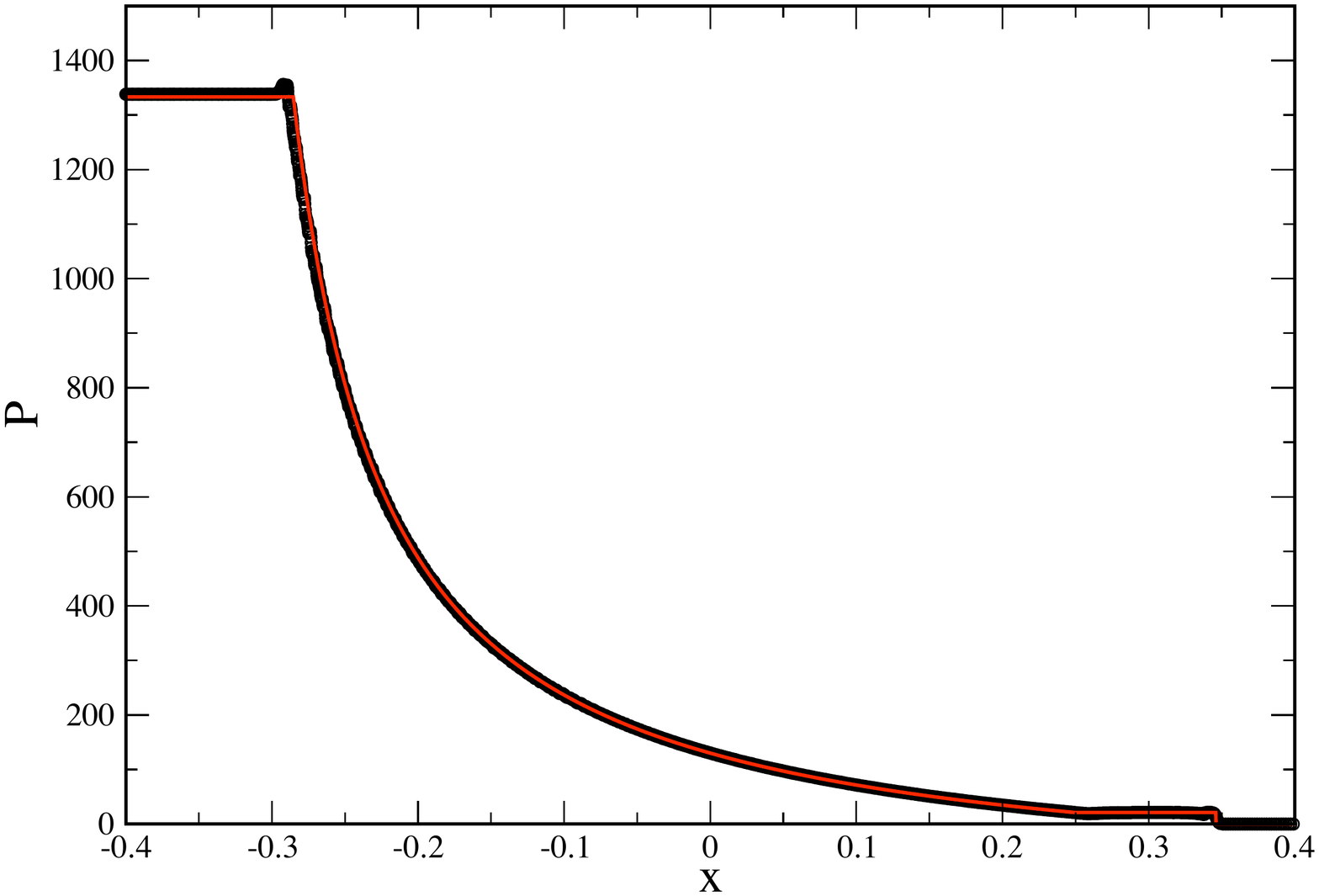}}
   \caption{Same as previous test, but the initial left hand side pressure has been increased 
            by a factor of 100. SPH results (at $t=0.35$) are shown as circles, the exact solution
            as red line. From left to right, top to bottom: velocity (in units of $c$), specific 
            energy, computing frame baryon number density and pressure.}
   \label{Rosswog::fig:3}
\end{figure}
This test is a more violent version of test 1 in which we increase the 
initial left side pressure by a factor of 100, but leave the other properties, in
particular the right-hand state,
unchanged: $[P, \rho, v]_{\rm L}= [4000/3, 10, 0]$ and $[P, \rho, v]_{\rm R}= [10^{-6},1,0]$.
This represents a challenging test since the post-shock density is compressed into a 
very narrow ``spike'', at $t=0.35$ near $x\approx 0.35$. A maximum Lorentz-factor of 
$\gamma_{\rm max} \approx 3.85$ is reached in this test.\\
In Fig.~\ref{Rosswog::fig:3} we show the SPH results (circles) of velocity $v$,
specific energy $u$, the computing frame number density $N$ and the pressure $P$ 
at $t= 0.35$ together with the exact solution of the problem (red line). Again
the numerical solution is in excellent agreement with the exact one, only in the specific
energy near the contact discontinuity occurs some smearing.

\subsection{Test 3: Riemann problem 3}
This test is an even more violent version of the previous tests. We now increase the 
initial left side pressure by a factor of 1000 with respect to test 1, but leave the other properties
unchanged: $[P, \rho, v]_{\rm L}= [40000/3, 10, 0]$ and $[P, \rho, v]_{\rm R}= [10^{-6},1,0]$.
The post-shock density is now compressed into a very narrow ``needle'' with a width of only $\approx 0.002$,
the maximum Lorentz factor is 6.65.\\
\begin{figure}[htbp] 
   \centering
 \centerline{
   \includegraphics[height=0.3\textheight]{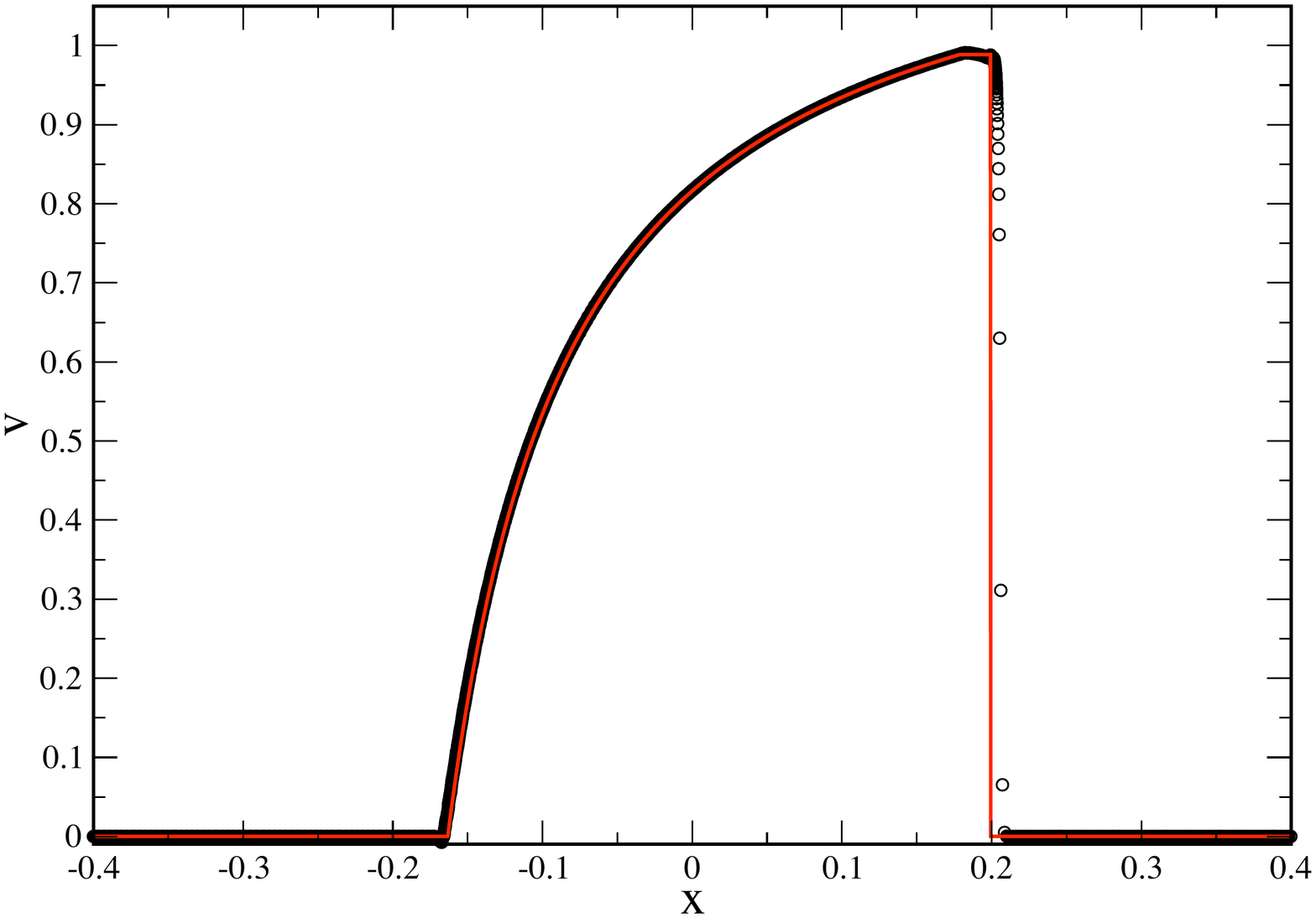} \hspace*{-0.8cm}
   \includegraphics[height=0.3\textheight]{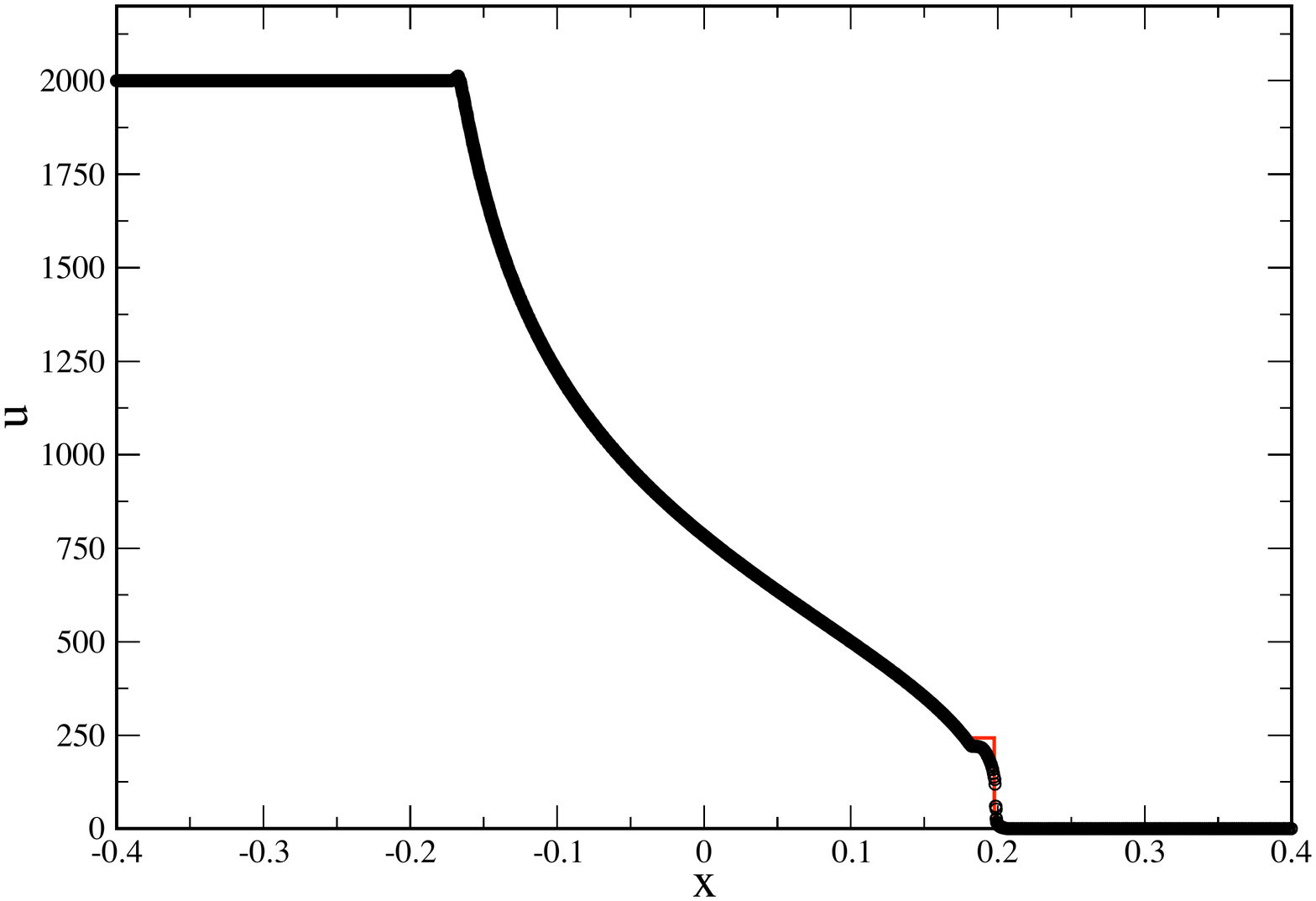}}
 \vspace*{-0.6cm}
   \centerline{
   \includegraphics[height=0.3\textheight]{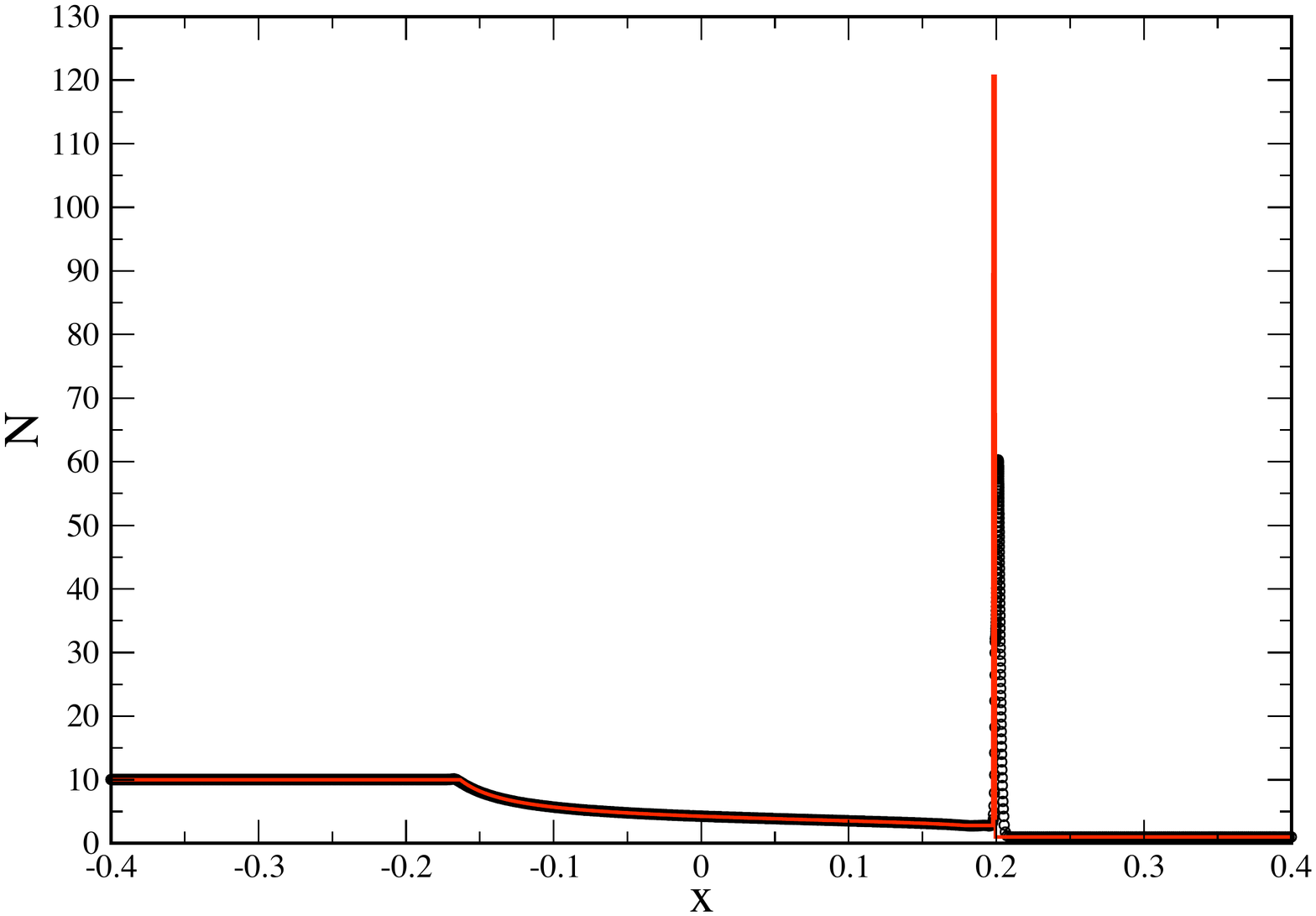} \hspace*{-0.8cm}
   \includegraphics[height=0.3\textheight]{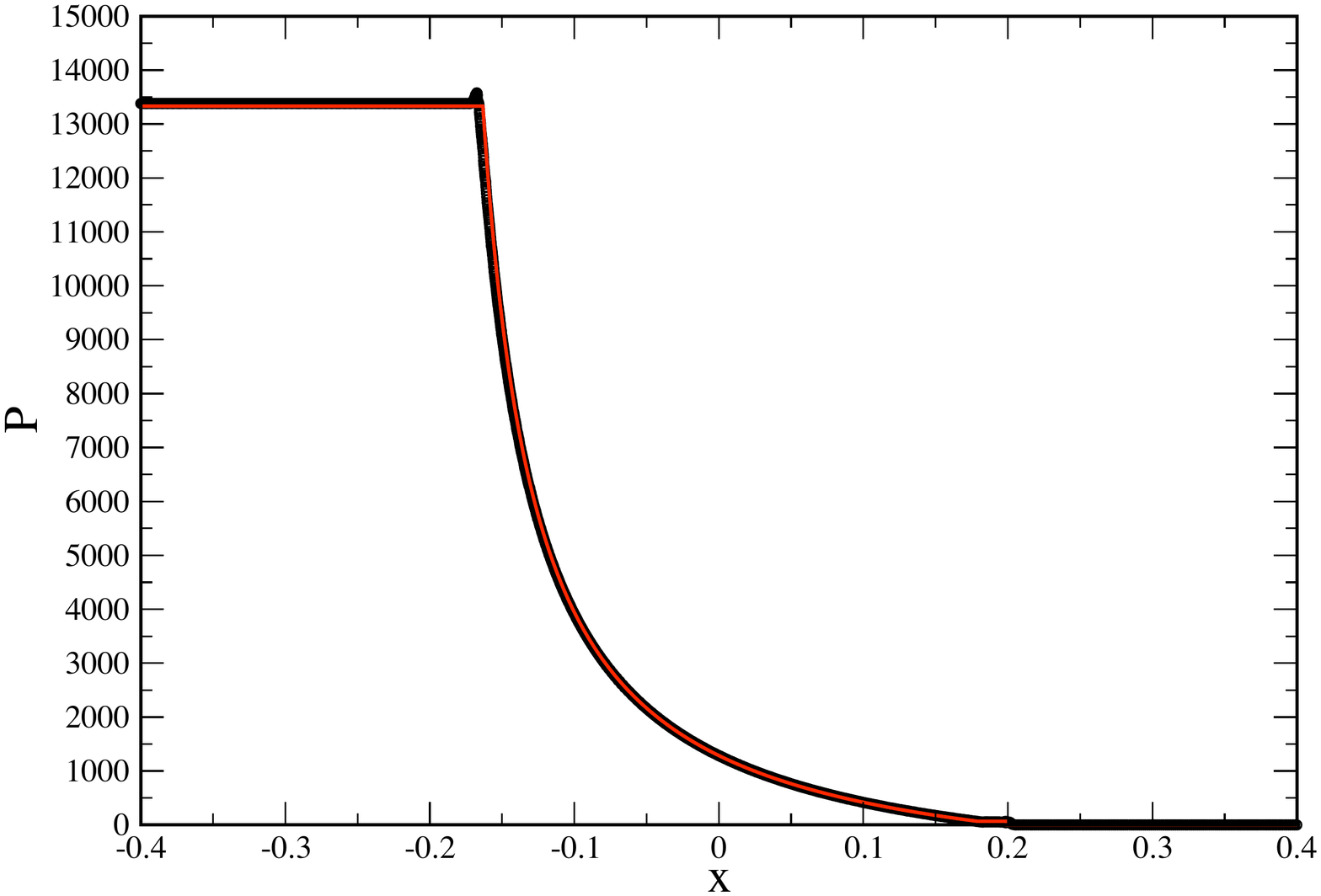}}
   \caption{Same as first shock tube test, but the initial left hand side pressure has been increased 
            by a factor of 1000. SPH results (at $t=0.2$) are shown as circles, 
            the exact solution
            as red line. From left to right, top to bottom: velocity (in units of $c$), specific 
            energy, computing frame baryon number density and pressure.}
   \label{Rosswog::fig:4}
\end{figure}
Fig.~\ref{Rosswog::fig:4} shows the SPH results (circles) of velocity $v$,
specific energy $u$, the computing frame number density $N$ and the pressure 
$P$ at $t= 0.2$ together with the exact solution (red line). The
overall performance in this extremely challenging test is still very good. The peak
velocity plateau with $v\approx 0.99$ (panel 1) is very well captured, practically no oscillations behind the shock
are visible. Of course, the ``needle-like'' appearance  of the compressed density shell 
(panel 3) poses a serious problem to every numerical scheme at finite resolution. At the applied
resolution, the numerical peak value of $N$ is only about half of the exact solution.
Moreover, this extremely demanding test  reveals an artifact of our scheme:
the shock front is propagating at slightly too large a speed. This problem decreases with 
increasing numerical resolution and experimenting with the parameter $K$ of 
Eqs.~(\ref{Rosswog::eq:diss_mom}) and (\ref{Rosswog::eq:diss_en}) shows that it is related 
to the form of artificial viscosity, smaller offsets occur for lower values of the viscosity 
parameter $K$. Here further improvements would be desirable.

\subsection{Test 4: Sinusoidally perturbed Riemann problem}
This is a more extreme version of the test suggested by \cite{Rosswog:dolezal95}. It starts from
an initial setup similar to a normal Riemann problem, but with the right state being sinusoidally
perturbed. What makes this test challenging is that the smooth structure (sine wave) needs to be transported
across the shock, i.e. kinetic energy needs to be dissipated into heat to avoid spurious post-shock oscillations,
but not too much since otherwise the (physical!) sine oscillations in the post-shock state are not accurately
captured. We use a polytropic exponent of $\Gamma=5/3$ and 
\be
 [P,N,v]^{\rm L}=[1000,5,0] \quad {\rm and} \quad [P,N,v]^{\rm R}=[5,2 + 0.3 \sin(50 x),0].
\ee
as initial conditions, i.e. we have increased the initial left pressure by a factor of 200 in comparison
to \cite{Rosswog:dolezal95}.
\begin{figure}[htbp] 
 \centering
   \centerline{
   \includegraphics[height=0.3\textheight,angle=0]{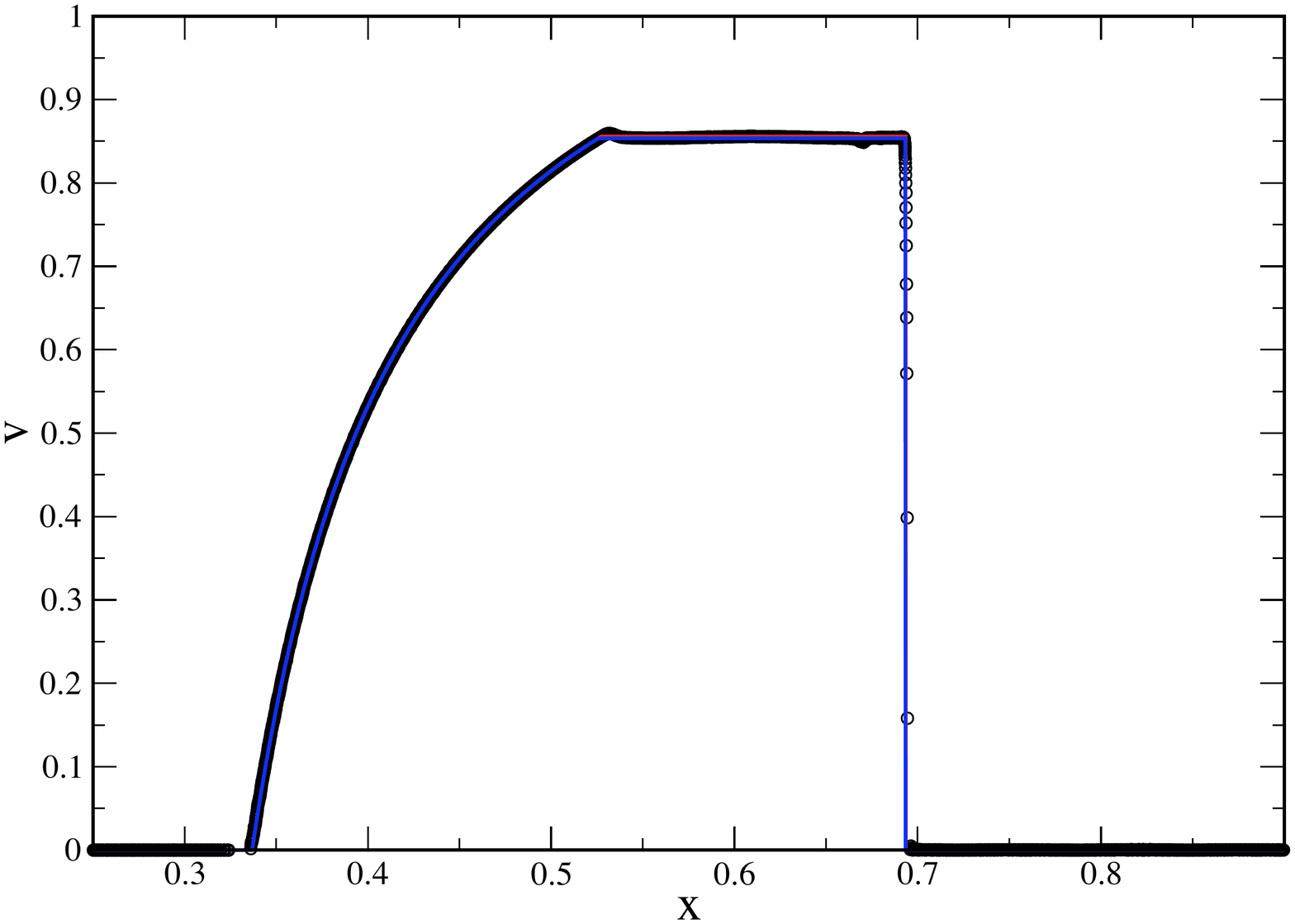} \hspace*{-0.6cm}
   \includegraphics[height=0.3\textheight,angle=0]{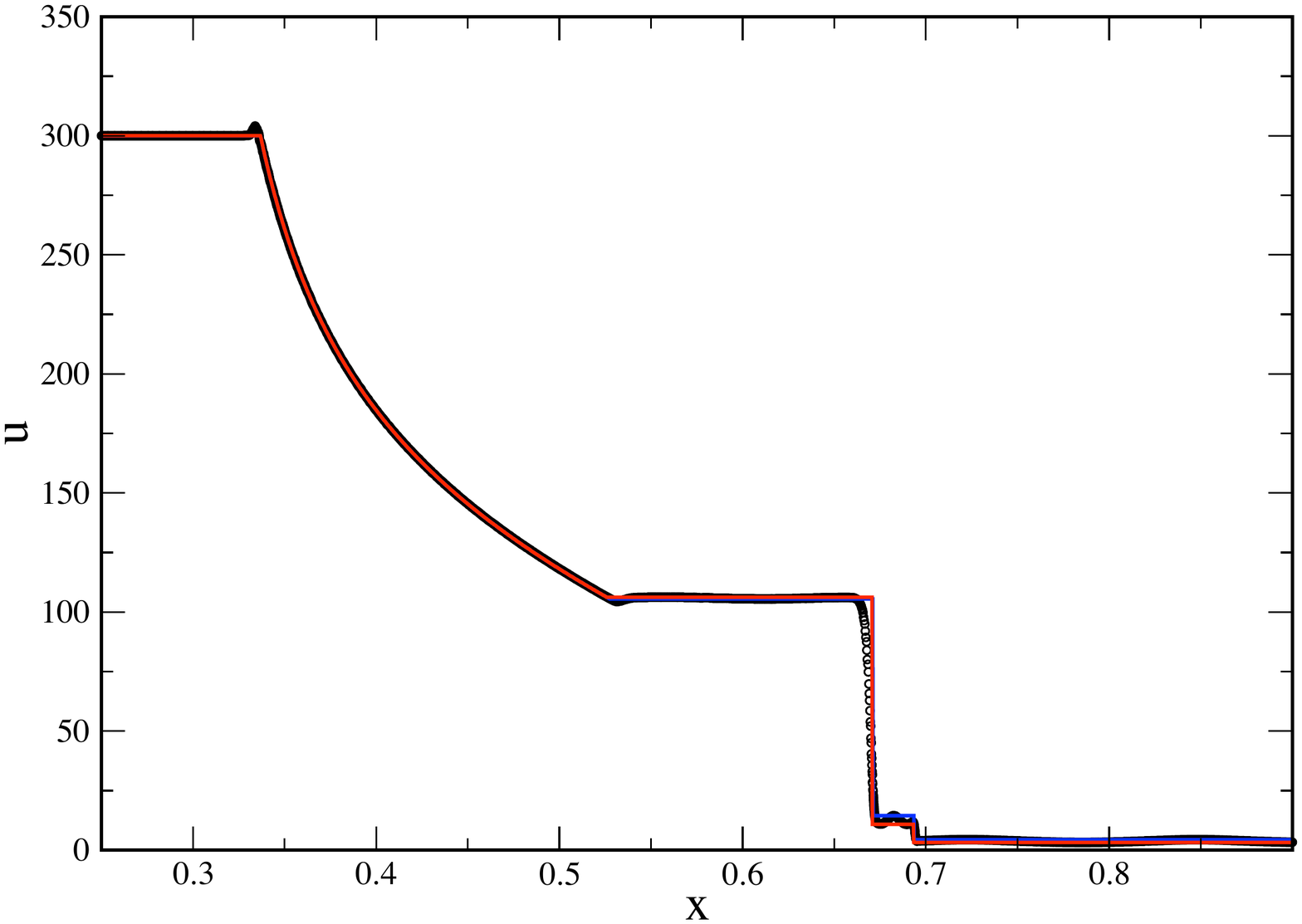} }
   \centerline{
   \includegraphics[height=0.3\textheight,angle=0]{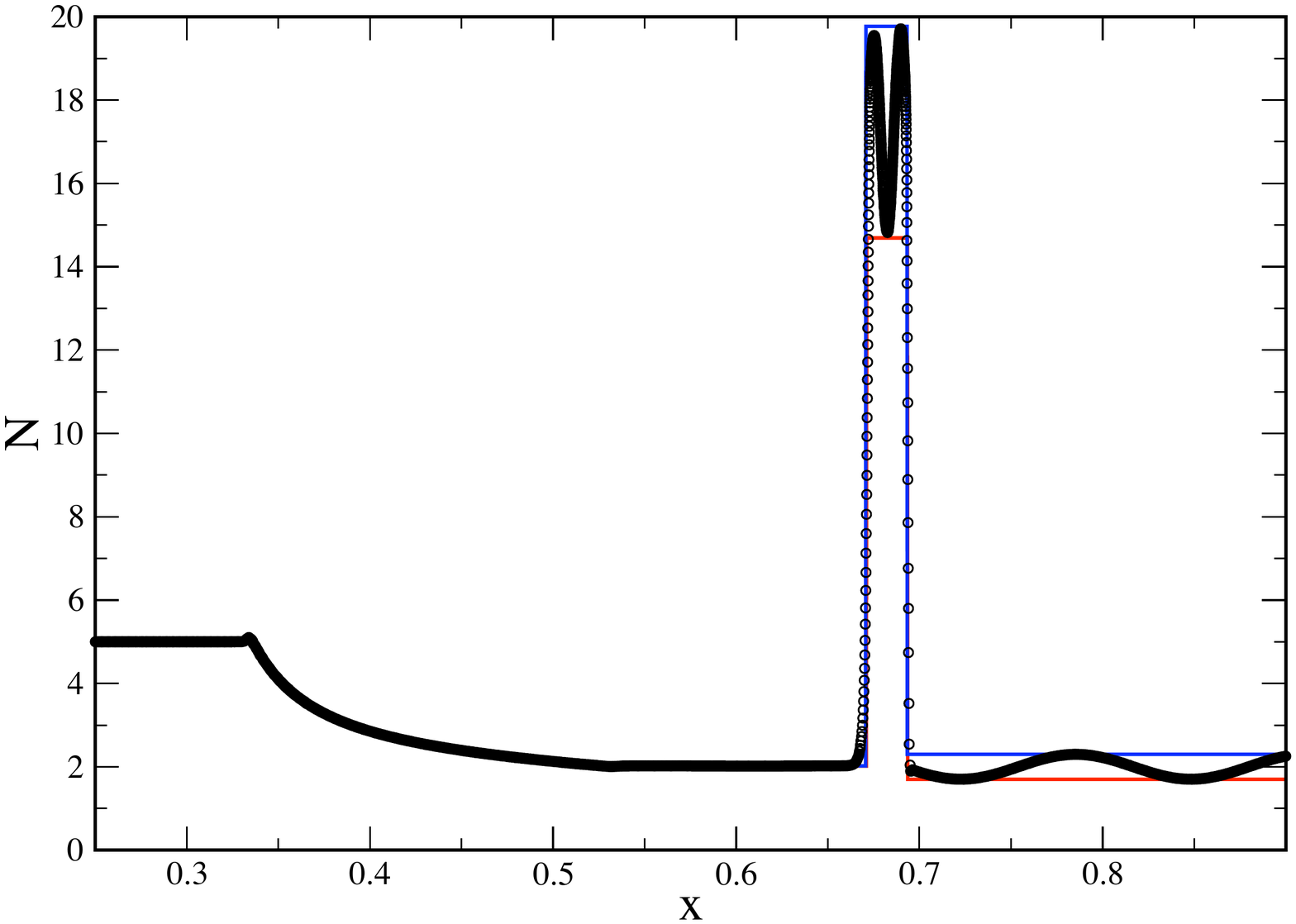} \hspace*{-0.6cm}
   \includegraphics[height=0.3\textheight,angle=0]{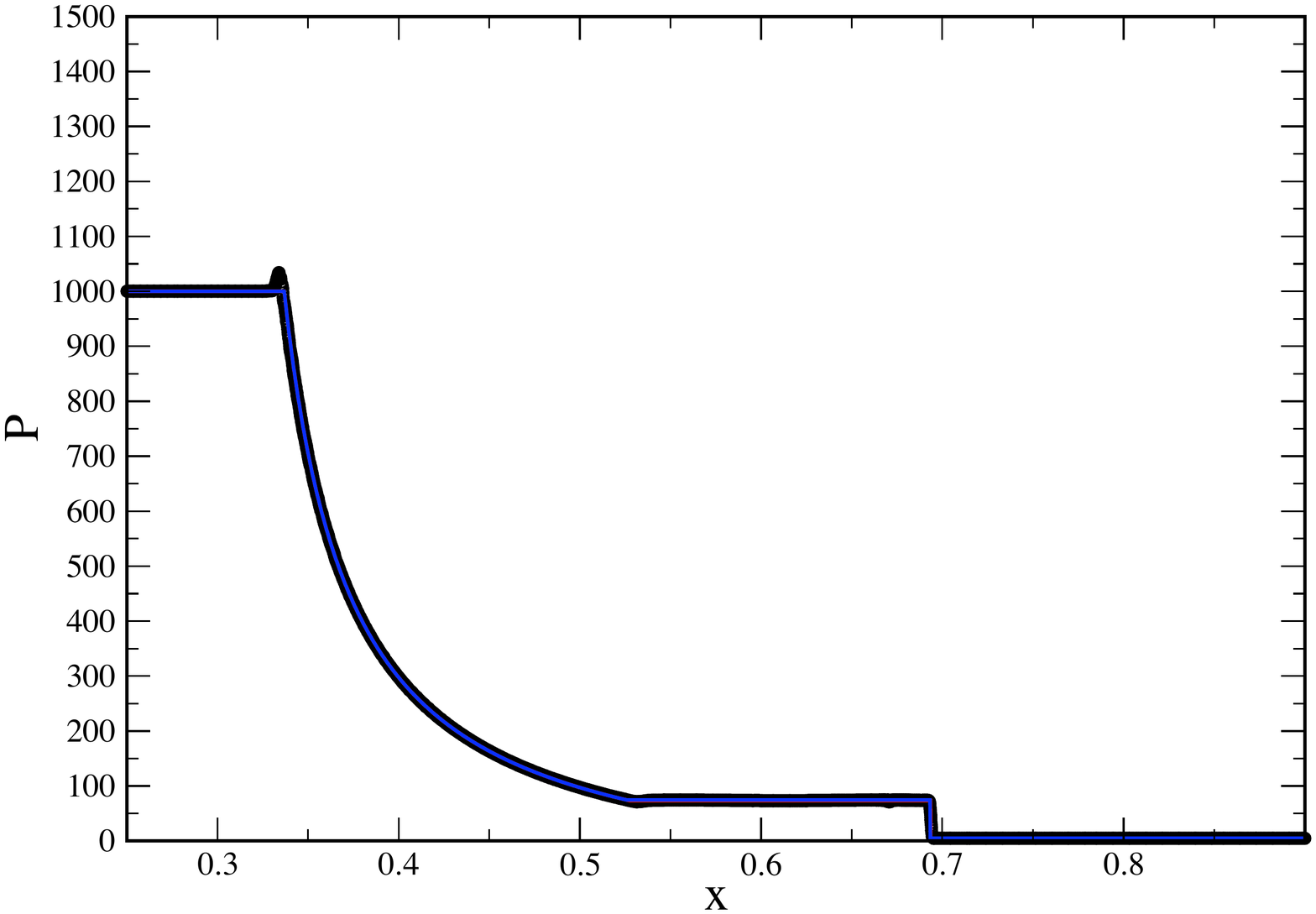} }
   \caption{Riemann problem where the right-hand side is periodically perturbed. The SPH solution
            is shown as circles, the exact solution for Riemann problems with constant RHS densities
            $N_{\rm R}= 2.3$ (blue) and $N_{\rm R}= 1.7$ (red) are overlaid as solid lines.}
   \label{Rosswog::fig:5}
\end{figure}
The numerical result (circles) is shown in Fig.~\ref{Rosswog::fig:5} together with two exact solutions,
for the right-hand side densities $N_{\rm R}=2.3$ (solid blue) and $N_{\rm R}=1.7$ (solid red). All the transitions
are located at the correct positions, in the post-shock density shell the solution nicely oscillates between the extremes
indicated by the solid lines.

\subsection{Test 5: Relativistic Einfeldt rarefaction test}
\begin{figure}[htbp] 
   \centering
   \centerline{
   \includegraphics[height=0.3\textheight,angle=0]{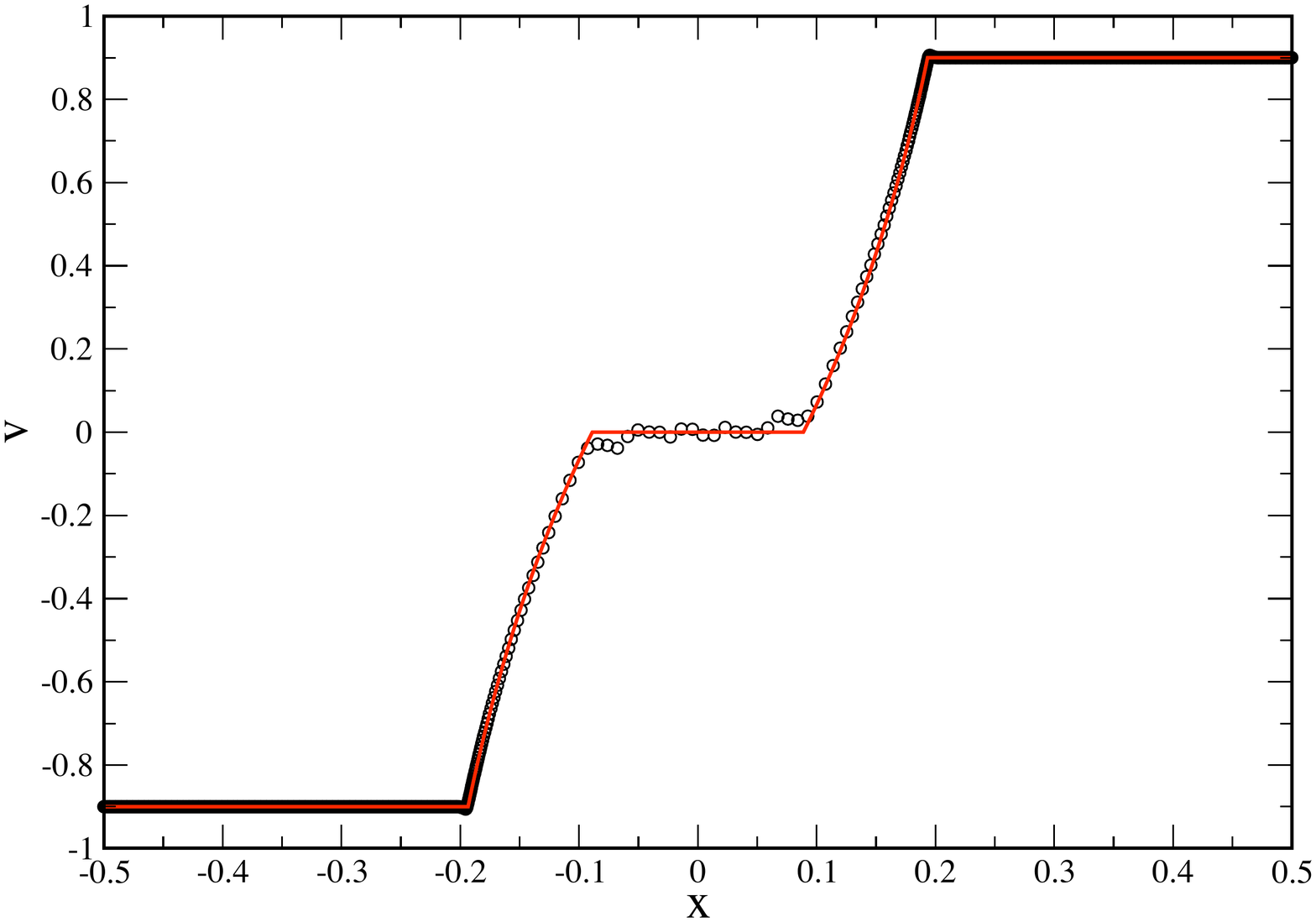} \hspace*{-0.6cm}
   \includegraphics[height=0.3\textheight,angle=0]{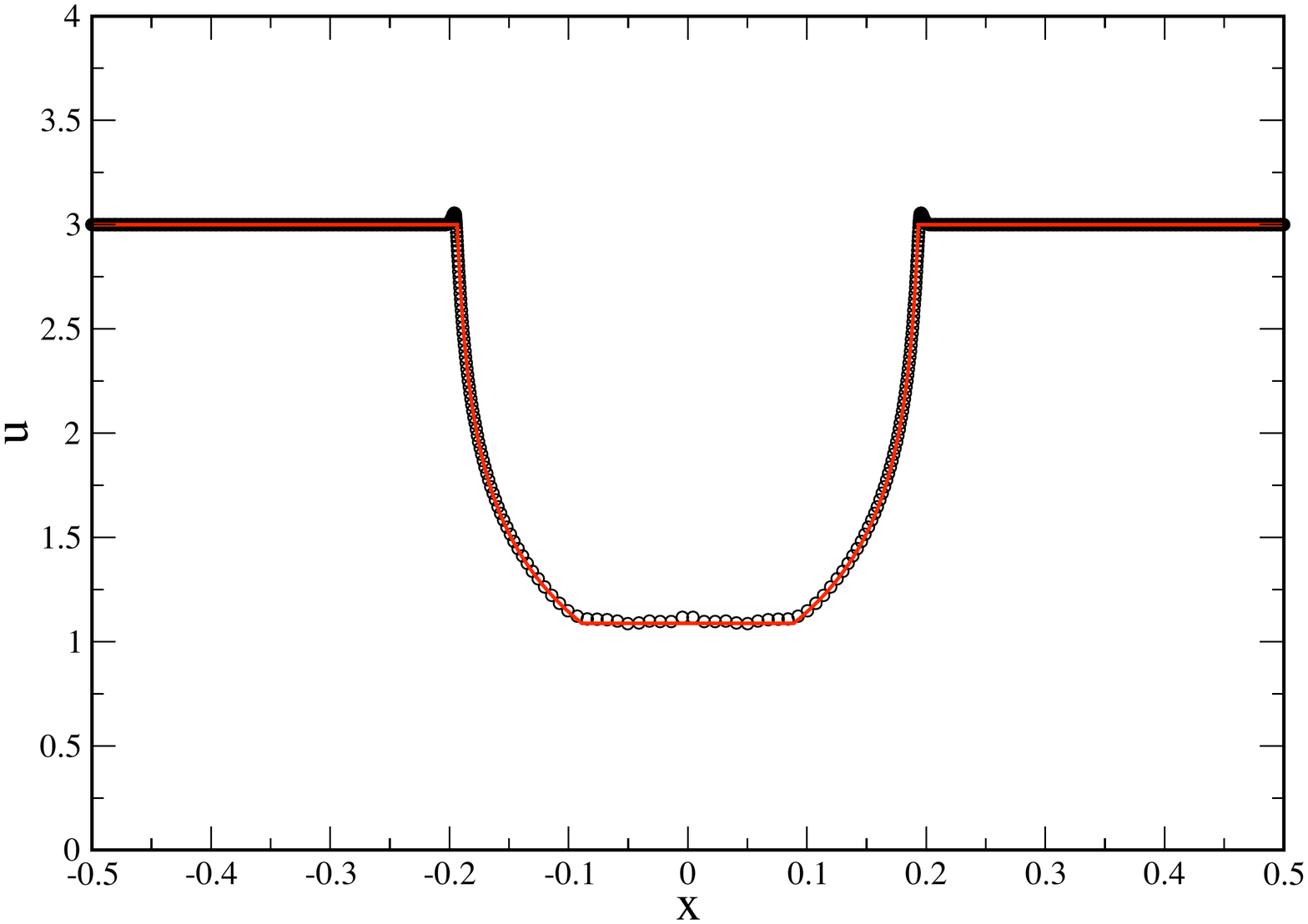} }
   \centerline{
   \includegraphics[height=0.3\textheight,angle=0]{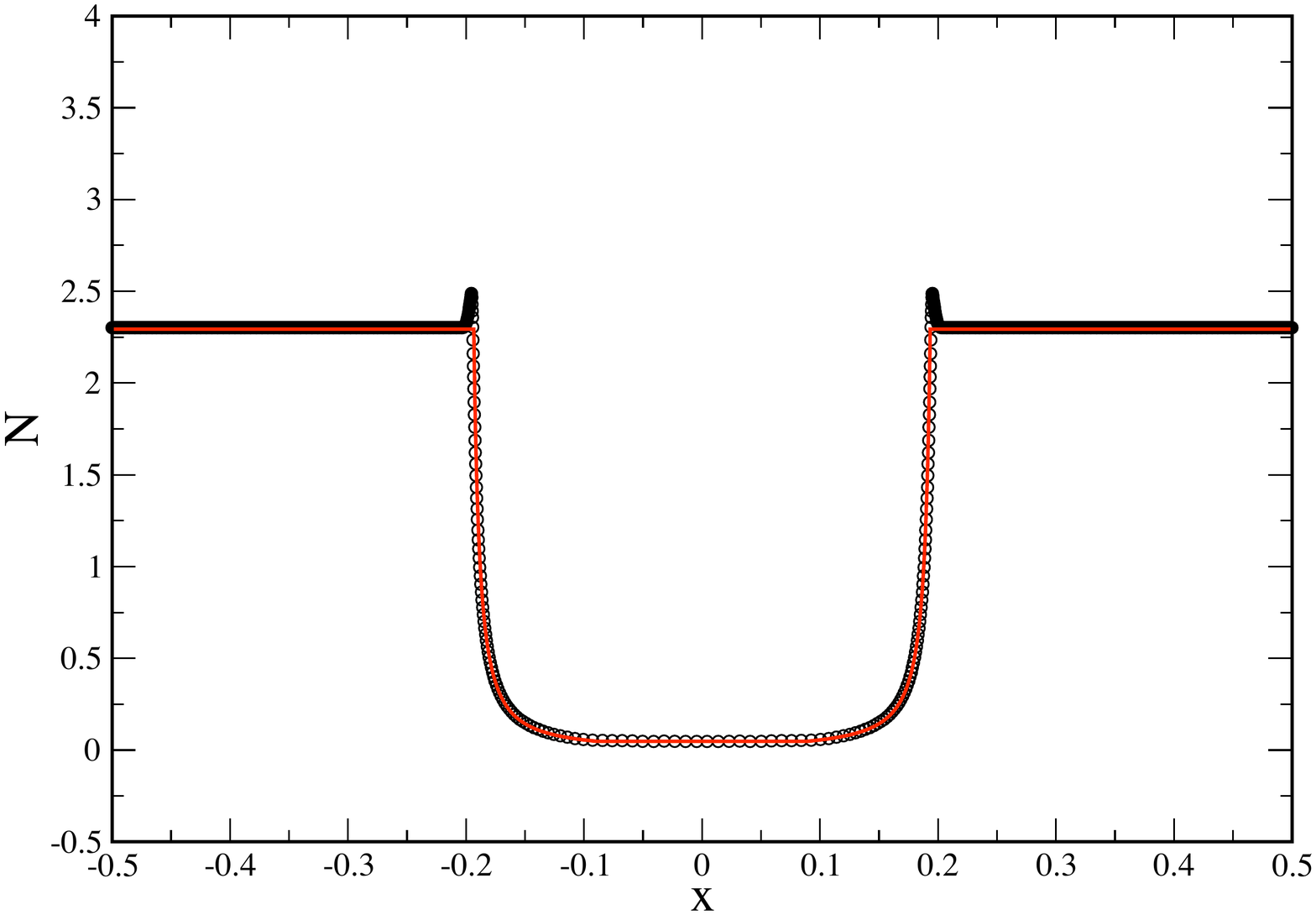} \hspace*{-0.6cm}
   \includegraphics[height=0.3\textheight,angle=0]{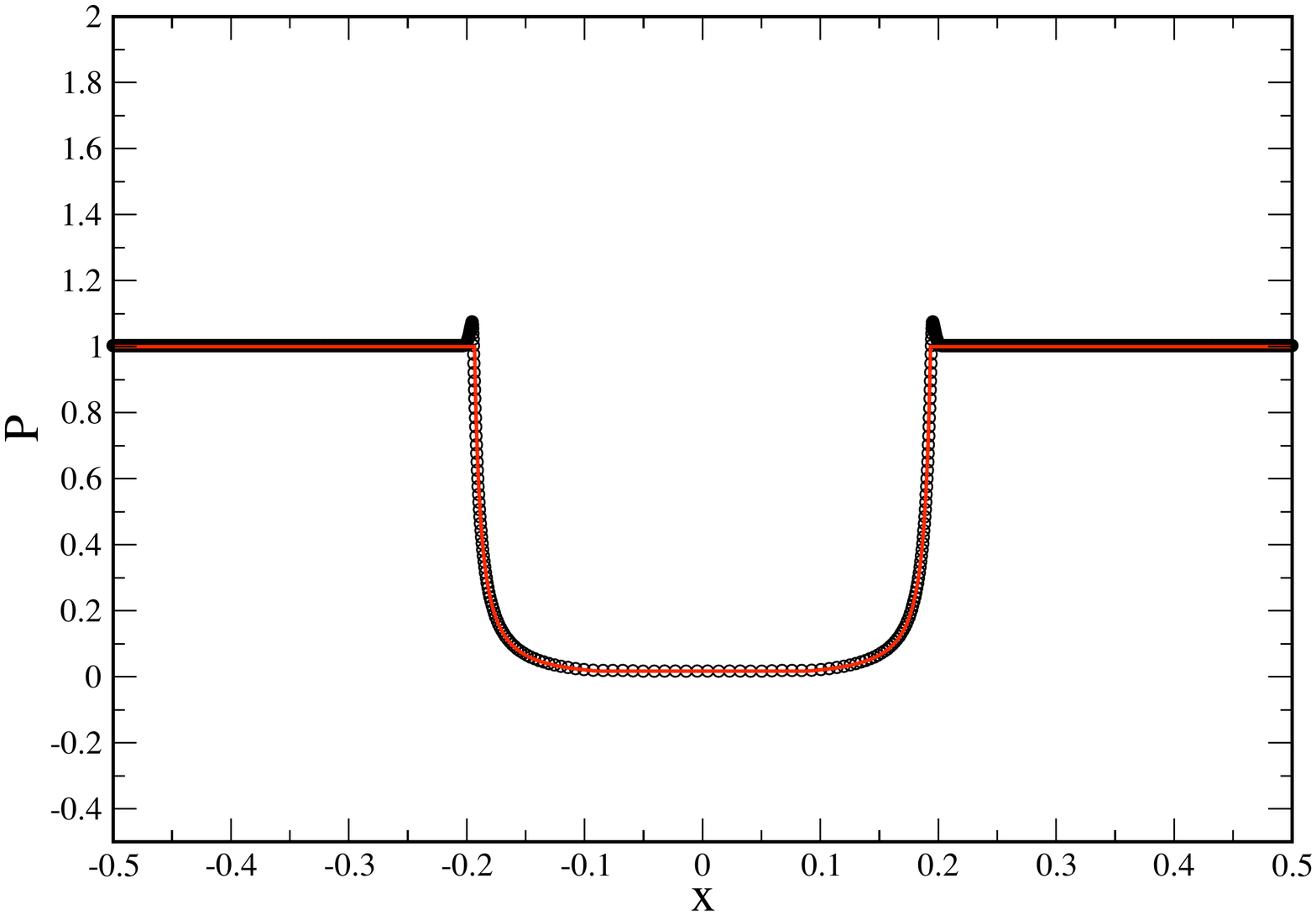} }
   \caption{Relativistic version of the Einfeldt rarefaction test. Initially the flow 
            has constant values of $n=1, P=1$ everywhere, $v_L= -0.9$ and $v_R=0.9$.}
   \label{Rosswog::fig:6}
\end{figure}

The initial conditions of the Einfeldt rarefaction test \cite{Rosswog:einfeldt91} do not exhibit discontinuities in
density or pressure, but the two halfs of the computational domain move in opposite directions and thereby
create a very low-density region around the initial velocity discontinuity. This low-density region poses
a serious challenge for some iterative Riemann solvers, which can return negative density/pressure values
in this region. Here we generalize the test to a relativistic problem in which left/right states move with 
velocity -0.9/+0.9 away from the central position. For the left and right state we use
$[P, n, v]_{\rm L}= [1, 1, -0.9]$ and $[P, n, v]_{\rm R}= [1, 1,  0.9]$ and an adiabatic 
exponent of $\Gamma=4/3$. Note that here we have specified the local rest frame density, $n$, which is related
to the computing frame density by Eq.~(\ref{Rosswog::eq:N_vs_n}). The SPH solution at $t=0.2$ is shown in 
Fig.~\ref{Rosswog::fig:6} as circles, the exact solution is indicated by the solid red line. Small oscillations 
are visible near the center, mainly in $v$ and $u$, and over-/undershoots occur near the edges of the rarefaction 
fan, but overall the numerical solution is very close to the analytical one. In its current form, the code can 
stably handle velocities up to 0.99999, i.e. Lorentz factors $\gamma > 200$, but at late times there are 
practically no more particles in the center (SPH's approximation to the emerging near-vacuum), so that it becomes
increasingly difficult to resolve the central velocity plateau.

\subsection{Test 6: Ultra-relativistic advection}
In this test problem we explore the ability to accurately advect a smooth density pattern
at an ultra-relativistic velocity across a periodic box. Since this test does not involve shocks
we do not apply any artificial dissipation. We use only 500 equidistantly placed particles in the 
interval $[0,1]$, enforce periodic boundary conditions and use a polytropic exponent of $\Gamma=4/3$.
We impose a computing frame number density $N(x)= N_0 + \frac{1}{2} \sin(2 \pi x)+\frac{1}{4} \sin(4 \pi x)$,
a constant velocity as large as  $v=0.99999999$, corresponding to a Lorentz factor of $\gamma \approx 7071$,
and instantiate a constant pressure corresponding to $P_0=(\Gamma-1)n_0 u_0$, where $n_0=N_0/\gamma$ and
$N_0=1$ and $u_0=1$. The specific energies are chosen so that each particle has the same pressure $P_0$.
With these initial conditions the specified density pattern should just be advected across the box without
being changed in shape.\\ 
\begin{figure}[htbp] 
   \centering
   \centerline{
     \includegraphics[height=0.38\textheight,angle=0]{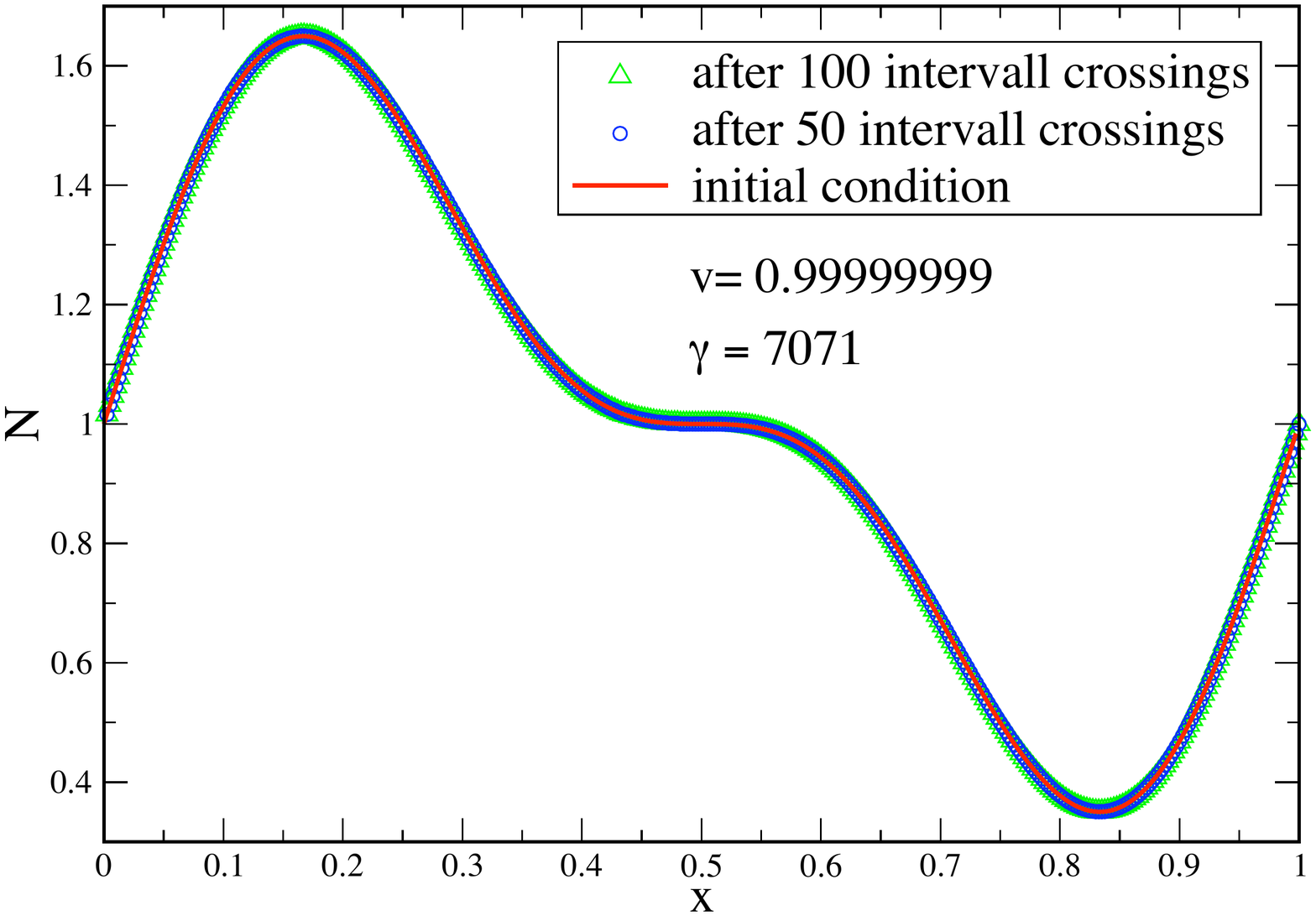} \hspace*{-1.cm}\includegraphics[height=0.38\textheight,angle=0]{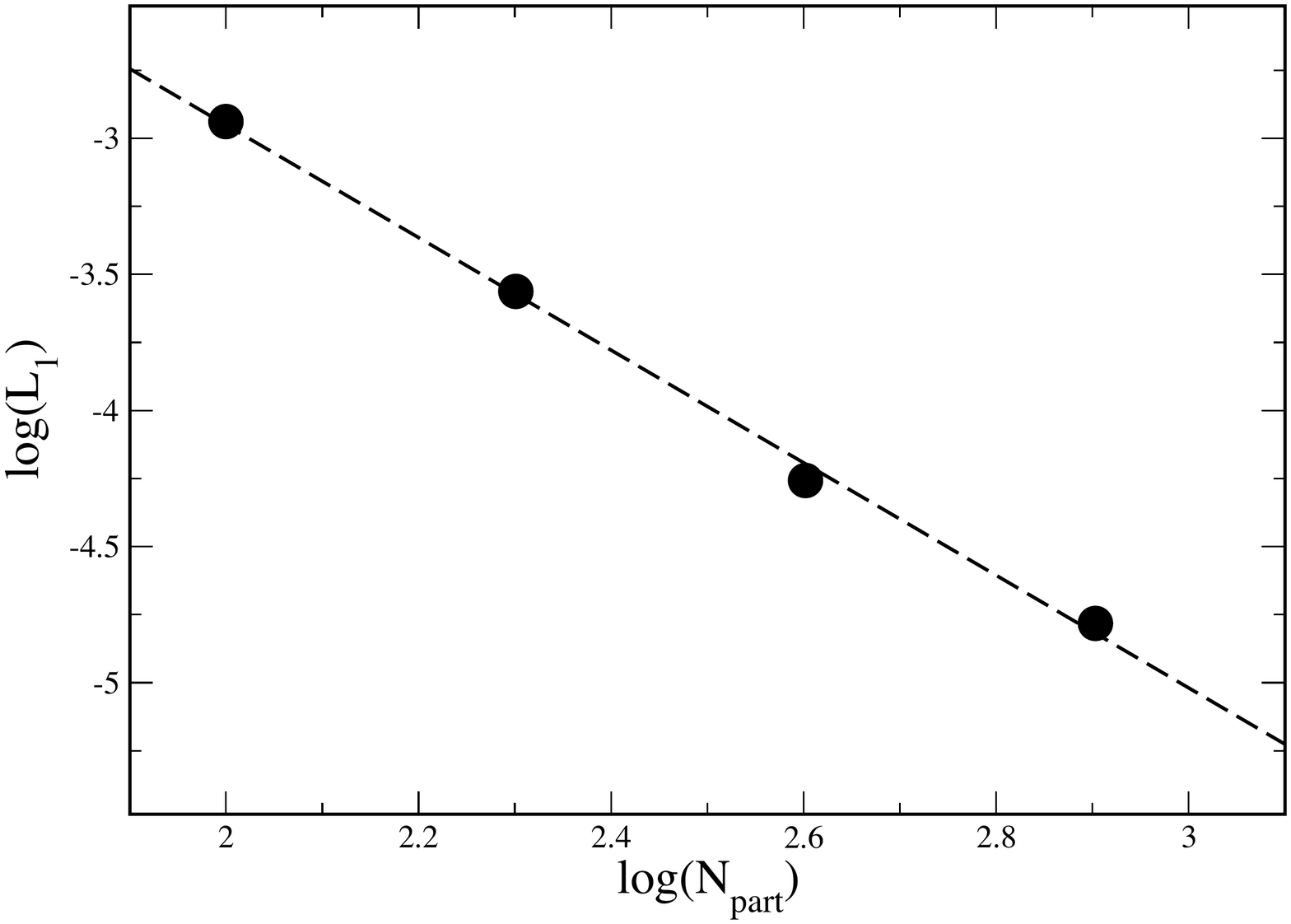}}
     \caption{Left: Ultra-relativistic advection ($v=0.99999999$, Lorentz factor $\gamma= 7071$) 
            of a density pattern across a periodic box. The advection is essentially perfect,
            the patterns after 50 (blue circles) and 100  (green triangles) times crossing the
            box are virtually identical to the initial condition (red line).
            Right: Decrease of the $L_1$ error as a function of resolution,
            for smooth flows the method is second-order accurate.}
   \label{Rosswog::fig:7}
\end{figure}
The numerical result after 50 times (blue circles) and 100 times (green triangles) crossing the interval 
is displayed in Fig.~\ref{Rosswog::fig:7}, left panel. The advection is essentially perfect, no deviation
from the initial condition (solid, red line) is visible.\\
We use this test to measure the convergence of the method in the case of smooth flow (for the case involving shocks,
see the discussion at the end of test 1). Since for this test the velocity is constant everywhere, we use the 
computing frame number density $N$ to calculate $L_1$ similar to Eq.~(\ref{eq:L1}). We find that the error decreases
very close to $L_1 \propto N^{-2}$, see Fig.~\ref{Rosswog::fig:7}, right panel, which is the behavior that is 
theoretically expected for smooth functions, the used kernel and perfectly distributed particles \cite{Rosswog:monaghan92} 
(actually, we find as a best-fit exponent -2.07). Therefore, we consider the method second-order accurate for smooth flows.

\section{Conclusions}
We have summarized a new special-relativistic SPH formulation that is derived from the Lagrangian of an
ideal fluid \cite{Rosswog:rosswog09d}. As numerical variables it uses the canonical energy and momentum per baryon whose
evolution equations follow stringently from the Euler-Lagrange equations. We have further applied the 
special-relativistic generalizations of the so-called ``grad-h-terms'' and a refined
artificial viscosity scheme with time dependent parameters.\\
The main focus of this paper is the presentation of a set of challenging benchmark tests that complement those
of the original paper \cite{Rosswog:rosswog09d}. They show the excellent advection properties of the method, 
but also its ability to accurately handle even very strong relativistic shocks. In the extreme shock tube
test 3, where the post-shock density shell is compressed into a width of only 0.1 \% of the computational
domain, we find the shock front to propagate at slightly too large a pace. This artifact ceases with 
increasing numerical resolution, but future improvements of this point would be desirable. We have further
determined the convergence rate of the method in numerical experiments and find it first-order accurate when
shocks are involved and second-order accurate for smooth flows.

\vspace*{0.5cm}

{\em Acknowledgment}\\
This work was supported by the German Research Foundation under grant number 50245 DFG-RO-5.

 \bibliographystyle{amsplain}


\end{document}